\documentclass[12pt]{elsarticle}
\pdfoutput=1
\usepackage{graphicx}

\usepackage{amssymb}
\usepackage{amsmath}
\usepackage{lineno}

\usepackage{color}
\usepackage{hyperref}
\usepackage[margin=3cm]{geometry} 
\usepackage[superscript,nomove]{cite}
\date{}

\begin{document}

\begin{frontmatter}

\title{Trip Centrality: walking on a temporal multiplex with non-instantaneous link travel time}

\author[1]{Silvia Zaoli}
\author[1]{Piero Mazzarisi}
\author[1]{Fabrizio Lillo}
\address[1]{Department of Mathematics, University of Bologna, Bologna, Italy}

\begin{abstract}

In complex networks, centrality metrics quantify the connectivity of nodes and identify the  most important ones in the transmission of signals. In many real world networks, especially in transportation systems, links are dynamic, i.e. their presence depends on time, and travelling between two nodes requires a non-vanishing time. Additionally, many networks are structured on several layers, representing, e.g.,  different transportation modes or service providers. Temporal generalisations of centrality metrics based on walk-counting, like Katz centrality, exist, however they do not account for non-zero link travel times and for the multiplex structure. We propose a generalisation of Katz centrality, termed Trip Centrality,  counting only the paths that can be travelled according to the network temporal structure, i.e. ``trips'', while also differentiating the contributions of inter- and intra-layer walks to centrality.   We show an application to the US air transport system, specifically computing airports' centrality losses due to delays in the flight network.\\

\end{abstract}
\end{frontmatter}
\thispagestyle{plain}


\section{Introduction}
\label{sec:intro}
Centrality metrics are a useful tool in network analysis to identify and rank the most important nodes or edges of a network. As different concepts of importance can be conceived, several centrality metrics have been proposed\cite{Newman2010}. The choice of the most suited one strongly depends on the application, on the type of network, on its structure and on how signals propagate on the network. 
Many centrality metrics are based on walks, paths, and distances on the network. Some consider only shortest paths and minimum distances, e.g. betweenness and closeness centrality, while others considers walks of all lengths. This is the case, e.g., of Katz centrality\cite{Katz1953}, PageRank\cite{Brin1998} and communicability\cite{Estrada2008, Estrada2012}, according to which a node's centrality is obtained by summing the contributions of all its outgoing (or incoming) walks, with longer walks being weighted less to reflect their smaller role in connecting the node to the rest of the network. The different metrics differ in the weighting scheme. The choice to consider walks of all lengths is motivated by the observation that processes taking place on the network, e.g. the diffusion of a signal, do not only use optimal paths\cite{Estrada2012}. Additionally, such choice permits a convenient computation of centrality based on the network's matrix description. \\
Many real world networks have a temporal structure\cite{Holme2012, Holme2015}, with links characterised by a time of appearance and a duration. In transportation networks, for example, a directed link between the departure and the arrival nodes appears at the time of departure and disappears at the time of arrival. Walks and paths on a temporal network must respect the temporal ordering of links, and this should be accounted for by temporal centrality metrics. While metrics for static networks can be applied to temporal networks after aggregating over time, i.e. considering all links to be present at the same time, clearly this procedure overestimates the number of walks and paths that can be traveled on the network and neglects the effect of the temporal dynamics on the network's connectivity\cite{Tang2009,Pan2011}. In fact, the links' temporal ordering would be completely disregarded. \textit{Ad hoc} centrality metrics for temporal networks are therefore needed, but how to perform this generalisation is not univocal. In fact, there might be different definitions of time-respecting walks and shortest paths (e.g., shortest according to the number of links used or to the time length), and different generalisation choices might be best suited for different applications. For example, different generalisations of betweenness and closeness centrality have been proposed \cite{Tang2010, Kim2012}, as well as other temporal metrics based on shortest paths and distances \cite{Pan2011, Kostakos2009, Holme2005}.\\
Metrics based on the counting of walks, like Katz centrality, are of particular interest in the temporal setting, because shortest paths might not be available at all times, therefore increasing the importance of longer itineraries. A natural way to generalise Katz centrality to the temporal case is to count only time-respecting walks. An extension based on this idea, called \textit{Dynamic communicability}, was proposed by Grindrod \textit{et al.}\cite{Grindrod2011}. However the definition of time-respecting walks used therein does not apply to cases where it takes a non-zero time to travel through a link. Yet, this is true for all transportation networks, for example, where the walk $i\rightarrow j \rightarrow k$ can be travelled only if the arrival in $j$ (disappearance of the $i\rightarrow j$ link) precedes the departure from $j$ to $k$ (appearance of the $j\rightarrow k$ link), and therefore link duration should be accounted for. It is also possible to generalise Katz centrality starting from its equivalent formulation, where the centrality of a node is proportional to the sum of the centralities of neighbouring nodes, plus a constant value. This choice has been done in Ref.[\citen{Taylor2017}], where each time interval is described by a different layer of the network, and each node is therefore a neighbour of itself at subsequent time steps. With this generalisation, however, the interpretation in terms of time-respecting walks is lost. \\
Additionally to the temporal structure, often networks comprise links of different types and are therefore better described as a multiplex, i.e. a network made of several layers, each containing (a copy of) all the nodes and only the links of a specific type. In transportation networks, for example, layers could represent different transportation services (e.g., bus, metro), service providers (e.g., bus operators or airlines) or lines of a transportation service. The importance of considering the multi-layer character of transportation network has been often stressed, e.g. to correctly characterise their topological properties \cite{Cardillo2013SR} or assess their resilience in response to failures \cite{Cardillo2013}. In the context of centrality metrics, this multiplex structure plays an important role because intra- and inter-layer walks might contribute differently to a node's centrality. For example, considering that flight connections between different airlines are riskier from the passenger's point of view, and therefore are probably less used, an airport should gain more centrality from an intra-layer walk of length $n$, using all flights of the same airline, rather than from an inter-layer walk of the same length. Centrality metrics for multiplex have been proposed, see e.g. Ref.[\citen{Boccaletti2014}] for a review. However, to our knowledge, none permit to weight differently inter- and intra-layer temporal walks. \\
Here, we propose a new centrality metric generalising Katz centrality to the case of a temporal multiplex with non-zero link travel time. Given the focus of this metric on counting only walks that can actually be travelled given the links schedule, we name it \textit{Trip Centrality}. First, in section \ref{sec:theory} we derive the generalisation starting from the static definition of Katz centrality in terms of walks on the network. Then, in section \ref{sec:comparison} we exemplify the fundamental differences between Trip Centrality and Dynamic communicability \cite{Grindrod2011} by applying both metrics to two toy networks with non-zero link travel time and showing the superior performance of Trip Centrality in this case. \\
A primary application of Trip Centrality is to transportation networks, and in particular we will consider air transport. The description of the air transport system as a network \cite{Zanin2013, Cook2015} proved useful to study its topological characteristics \cite{Guimera2005,Cardillo2013SR}, resilience \cite{Cardillo2013,Verma2014}, epidemic spreading \cite{Colizza2006}, and delay propagation \cite{Fleurquin2013}. Delays change the timing of links, and their effects in terms of missed connections (and consequently of costs for airlines) depend in a complex way on the schedule. In section \ref{sec:Data} we apply Trip Centrality to the US air transport network, and by comparing the network of scheduled and realised flights we show the capability of the new metric to identify the airports' loss of centrality due to delays, something that is impossible with the static metrics and with existing temporal metrics not accounting for the non-zero link travel time. The loss of centrality of an airport quantifies the itineraries connecting that airport to the rest of the network that become unfeasible due to delays, and therefore assess the effects of delays at the network level. 

\section{Results}
\label{sec:results}
\subsection{Definition of Trip Centrality}
\label{sec:theory}
Consider a static directed network of $N$ nodes with weighted adjacency matrix $A_{ij}$, such that $A_{ij}=k$ if there are $k$ links from $i$ to $j$ (for example, flights). The outgoing (incoming) Katz centrality of node $i$ is given by the sum of the contributions of walks of any length outgoing from (incoming to) $i$, where each walk of length $n$ contributes $\alpha^n$. Then, given that $A^n_{ij}$ is the number of walks of length $n$ from $i$ to $j$:
\begin{subequations}
\label{eq:katz}
 \begin{align}
 k_i^{out}&=\sum_{n=0}^\infty \alpha^n \sum_{j=1}^N (A^n)_{ij}=\sum_{j=1}^N \bigl[\sum_{n=0}^\infty (\alpha A)^n\bigr]_{ij}=\sum_{j=1}^N [(\mathbb{I}-\alpha A)^{-1}]_{ij},\\
  k_i^{in}&=\sum_{n=0}^\infty \alpha^n \sum_{j=1}^N (A^n)_{ji}=\sum_{j=1}^N \bigl[\sum_{n=0}^\infty (\alpha A)^n\bigr]_{ji}=\sum_{j=1}^N [(\mathbb{I}-\alpha A)^{-1}]_{ji},\\
 \end{align}
\end{subequations}

\begin{figure}[t]
\begin{center}
\includegraphics{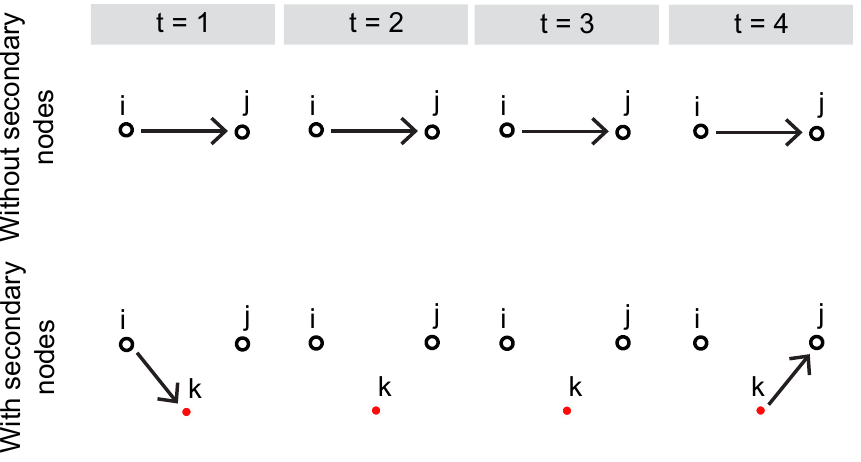}
\caption{Example showing how secondary nodes are introduced. Above, a link between the primary nodes $i$ and $j$, appearing during time frame 1 and disappearing at time frame 4. Below, the same link is represented by two stubs, one at time frame 1 between $i$ and the secondary node $k$, and one at time frame 4 between $k$ and $j$. }
\label{fig:secondary_nodes}
\end{center}
\end{figure}

Note that the infinite sums on $n$ in Eqs. \ref{eq:katz} converge only if $\alpha$ is smaller than the inverse of the largest eigenvalue of $A$ .\cite{Newman2010}\\
In a temporal network, links are characterised by their time of appearance and their duration, i.e. the length of the time interval during which they are present. Here, we will take the duration to coincide with the time required for a signal to travel through the link. This assumption is inspired by transportation networks, where a link represents a route of a transportation service, appearing at the departure time and disappearing at the arrival time.  A common way to treat temporal networks is to discretise time \cite{Grindrod2011, Taylor2017} in intervals of length $\Delta \tau$ and define an adjacency matrix $A^{[t]}$ for each time frame. The adjacency matrix $A^{[t]}$ contains only the links that either appear or disappear in the interval $[\tau,\tau+\Delta \tau)$, or are present during the entire interval. With this choice, the temporal network is represented by a series of $T$ adjacency matrices $\{A^{[t]}\}_{t=1,...,T}$. The optimal length of a time frame $\Delta \tau$ depends on the dataset, as will be discussed in the following. \\
In the case of non-zero link travel times, however, the above formulation is not suited to count time-ordered walks. We define a time-ordered walk as a series of links such that the disappearance of a link in the series always precedes the appearance of the following link. In the air transport example, a time-ordered walk is a series of flights such that the $i$-th lands before the $(i+1)$-th departs. Products $A^{[t_1]}\dots A^{[t_n]}$ of adjacency matrices defined as above would count also walks such that the $(i+1)$-th links appears when the $i$-th is still present. 
In order to express the number of time-ordered walks as a product of adjacency matrices, we introduce a set of secondary nodes, one for each of the $N_l$ links present over the whole considered period (see Fig.\ref{fig:secondary_nodes}). We therefore consider a network with $N+N_l$ nodes, of which $N$ are the original, or primary, nodes and $N_l$ are the secondary ones. In this new network, a link from $i$ to $j$, appearing during time frame $t$ and disappearing during time frame  $t'>t$, is associated with a secondary node $k$ and split into two links, called `stubs' in the following, one from $i$ to $k$  present during time frame $t$, and one from $k$ to $j$  present during time frame $t'$. We remark that, differently from links, stubs do not have a duration, i.e. they exist only in one time frame, and in the time frames between the appearance and disappearance of a link no stub related to that link is present in the network. For each time frame $[\tau,\tau+\Delta \tau)$ we define an adjacency matrix $A^{[t]}$ of size $(N+N_l)\times(N+N_l)$ such that $A_{ij}^{[t]}=1$ either if a link outgoing from node $i$ appears during that time frame, and $j$ is the secondary node associated with it, or if a link incoming to node $j$ disappears during that time step, and $i$ is the the secondary node associated with it. In the air transport example, for each flight a directed stub between the origin airport and its secondary node is present in the time frame in which the flight's departure time falls, while a directed stub between its secondary node and the destination airport is present in the time frame in which the arrival time falls. Secondary nodes ensure that matrix products of the form $A^{[t_1]}A^{[t_2]}\dots A^{[t_n]}$, with $t_1<t_2<\dots <t_n$ (thus, without repetitions, meaning that at most one link is used per time frame) count only time-ordered walks, in the sense defined above. The time frames $t_1,\dots, t_n$ do not need to be consecutive, as a walk can pause at a node for some time frames before continuing. We note that the idea of secondary nodes to describe temporal networks with non-instantaneous transport has been suggested previously \cite{Kempe2002, Zanin2009}, but never applied to the computation of centrality metrics. The length $\Delta \tau$ of one time frame must be shorter than the duration of the shortest link, so that the two stubs in which the link is split belong to different time frames and can be both used in a walk. With this definition of the series of adjacency matrices, the $i,j$-th element of the matrix
\begin{equation}
\label{eq:Q}
    Q=[(\mathbb{I}+\tilde{\alpha} A^{[1]})(\mathbb{I}+\tilde{\alpha}  A^{[2]})\dots(\mathbb{I}+\tilde{\alpha}  A^{[T]})-\mathbb{I}],
\end{equation}
contains the contribution to centrality of all walks from $i$ to $j$. In fact, $Q$ contains all the time-ordered products of $n$ adjacency matrices, for $n=1,...,T$. Note that $\tilde{\alpha}$ is the weight of a one-stub walk, therefore a one-link walk, which uses two stubs, is weighted $\alpha=\tilde{\alpha}^2$. The vectors of temporally generalised outgoing and incoming Katz centrality are then obtained summing $Q$, respectively, over columns and rows:
 \begin{subequations}
 \label{eq:trip_S}
 \begin{align}
    \vec{t}_S^{out}&=Q\vec{1}_{N+N_l},\\
    \vec{t}_S^{in}&=\vec{1}_{N+N_l}^T Q,
    \end{align}
\end{subequations}
where $\vec{1}_{N+N_l}$ is a column vector of ones and the subscript $S$ indicates that this is a single layer quantity. We refer to these centralities as \textit{Single-Layer Trip centralities}. Note that, differently from the static case, there is no upper limit for the parameter $\tilde{\alpha}$, as the sum is always bounded. In fact, there are no walks longer than $T$.\\

Let us now consider the multiplex structure of the temporal network. We associate one layer with each type of link existing in the network. The purpose of this further generalisation is to distinguish the walks made of links of the same type from those made of links of different types, which might give a different contribution to the centrality (length being equal). As the multiplex has a copy of each node on each layer, the adjacency matrix $A$ is of size $(N N_L+N_l)\times(N N_L+N_l)$, where $N_L$ is the number of layers. For each link of type $\lambda$ a directed stub is present, in the time frame corresponding to its appearance, between the copy of the origin node on layer $\lambda$ and its secondary node (laying on the same layer). A directed stub between its secondary node and the copy of the destination node on layer $\lambda$ is present in the time frame corresponding to its disappearance. \\
Let us introduce the parameter $\varepsilon\le1$, such that the contribution to centrality of a walk of length $n$ changing layer $m$ times is $\varepsilon^m \alpha^n$. The effect of this parameter is that, the more changes of layer a walk has, the less it contributes to centrality. Let us then introduce the matrix $K$, of the same size of $A$, as the matrix with elements $K_{ii}=1$ and $K_{ij}=\varepsilon$ if $i$ and $j$ are two copies of the same node on different layers. Now, the products of the form $A^{[t_1]}KA^{[t_2]}K\dots K A^{[t_n]}$ count walks by introducing a factor $\varepsilon$ every time there is a change of layer. Therefore, the outgoing Trip Centrality  on the temporal multiplex is written as

\begin{equation}
\label{eq:Trip}
  \vec{t}^{out}=[(\mathbb{I}+\tilde{\alpha} A^{[1]}K)(\mathbb{I}+\tilde{\alpha}  A^{[2]}K)\dots(\mathbb{I}+\tilde{\alpha}  A^{[T]}K)-\mathbb{I}]K^{-1}\vec{1}_{(N N_L+N_l)}.
\end{equation}
The incoming centrality is generalised similarly.\\
With this procedure, we obtain a centrality measure for each copy of each node, which sums the contributions of walks outgoing from (or incoming to) that node with a link on the layer on which that copy lies. Such \textit{layer-specific centrality} is interesting if we want to measure the importance of a node for, e.g., a specific transportation service or service provider. However, if we are interested in the role of the node for the entire network, an aggregated measure is needed. The interpretation of centrality in terms of walks provides us with a natural way to perform such an aggregation. In fact, if we compute the \textit{aggregated centrality} of a node by summing the centralities of all its copies, we obtain the contribution of all walks outgoing from (or incoming to) any copy of that node. Additionally, the centrality of a secondary node represents the centrality of the corresponding link. Given that links between the same two nodes having different schedules (e.g., flights between the same two airports at different times of the day) are associated with different secondary nodes, the centrality of their secondary nodes can show how the importance of a link changes depending on its schedule. \\
The parameter $\varepsilon$ determines how much inter-layer walks are penalised. In the case of transportation networks, therefore, it measures the propensity of users to change layer. In particular, the case $\varepsilon =0$ corresponds to not allowing inter-layer walks. In this case, the layer-specific centralities are those that consider only the links of one type. The aggregated centralities consider links of all layers, but only intra-layer walks contribute. When instead $\varepsilon=1$ , inter-layer walk give the same contribution of intra-layer walks. For $0<\varepsilon<1$, inter-layer walks are considered, but contribute less than an intra-layer walk of the same length.  Note that the matrix $K$ is invertible for every $\varepsilon \neq 1$. The final multiplication by $K^{-1}$ in equation \eqref{eq:Trip} only changes the aggregated centralities by a multiplicative factor, therefore it does not change the node's ranking (see SI, section 1). Thus, the ranking according to the aggregated centralities in the case $\varepsilon=1$ can be obtained skipping the final multiplication by $K^{-1}$. \\
A generalisation along the same lines is also possible for PageRank, a centrality metrics developed by Google \cite{Brin1998} which, similarly to Katz centrality, can be defined in terms of walks. In PageRank, an additional weight is assigned to the walks, depending on the in- (or out-) degree of the nodes they cross. See section 2 of the SI for details. 

\subsection{Comparison with Dynamic communicability}
\label{sec:comparison}
\begin{figure}[t]
\begin{center}
\makebox[\textwidth][c]{\includegraphics{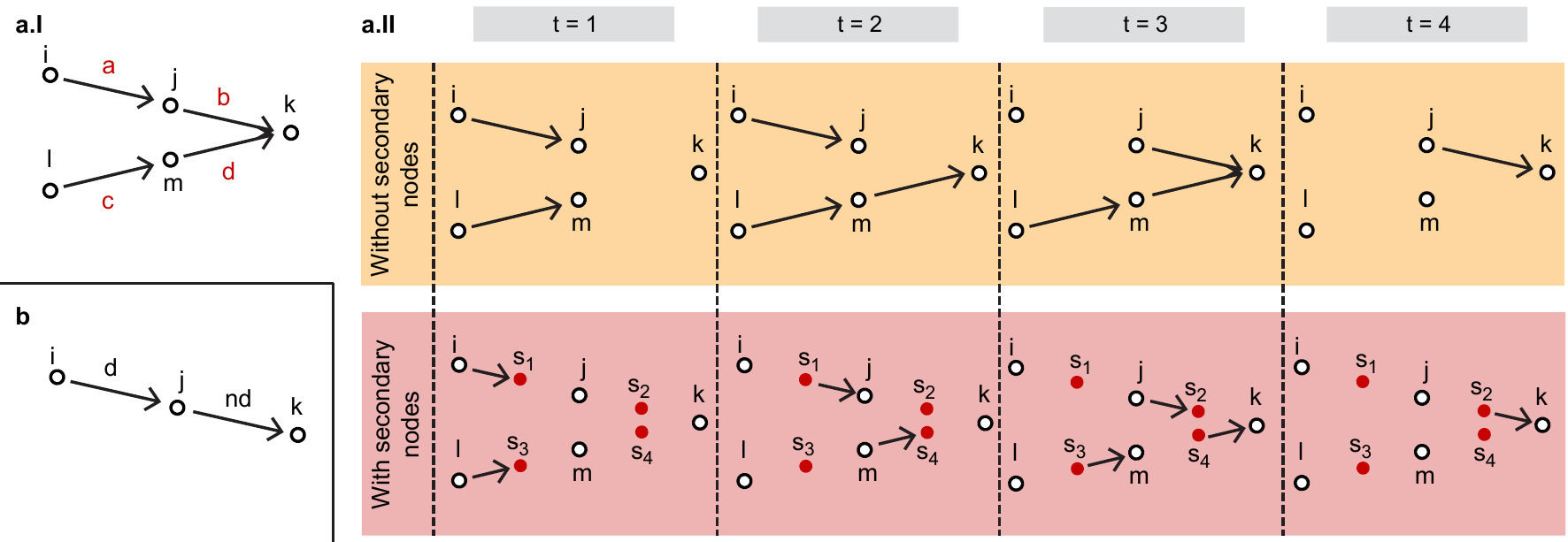}}%
\caption{Two examples illustrating the difference between Trip Centrality and Dynamic communicability in ranking nodes on temporal networks with non-zero link travel time.  a.I) Time-aggregated representation of a single-layer temporal network. Link $a$ is present during the time frames 1 and 2, link $b$ during the time frames 3 and 4, link $c$ from time frame 1 to time frame 3, and link $d$ during the time frames 2 and 3; a.II) Time-explicit representation of the network in panel a.I. Above, the representation without the use of secondary nodes, below with; b) Time-aggregated representation of a single-layer temporal network. The link from $i$ to $j$ has a duration $d$, the link from $j$ to $k$ has a duration $nd$ and appears after an interval $md$ from the disappearance of the former. }
\label{fig:examples}
\end{center}
\end{figure}

In this section, we show the difference between the proposed Trip Centrality and Dynamic communicability, a previously proposed temporal generalization of Katz centrality \cite{Grindrod2011}, by applying them to two simple temporal networks. The purpose of the two examples is to show that, in the case of non-zero link travel time, Trip Centrality overcomes some limitations of Dynamic communicability related to time-respecting walks and to the time-discretisation. Note that two versions of Dynamic communicability are presented in Ref. [\citen{Grindrod2011}], one allowing multiple jumps during each time frame, and one allowing only one, similarly to Trip Centrality. We consider the second version for the comparison. \\
First, we consider the network in figure \ref{fig:examples}a.I, consisting of four nodes, all on the same layer. The links' temporal dynamics, represented in figure \ref{fig:examples}a.II, is such that it is possible to move from $i$ to $k$ respecting the schedule, but not from $l$ to $k$. In fact, link $c$ from $l$ to $m$ disappears only after the appearance of link $d$ from $m$ to $k$. In a network where the link duration coincides with the link travel time, this means that the connection between links $c$ and $d$ cannot be taken. Instead, it is possible to use links $a$ and $b$ in sequence to go from $i$ to $k$. 
Therefore, the outgoing centrality of node $i$ should be larger than the one of $l$, as it has one outgoing itinerary using one link and one using two links, while $k$ has only one outgoing itinerary, using one link. However, if we compute the outgoing centralities of $i$ and $l$ according to Dynamic communicability, which means not introducing secondary nodes, $l$ is found to be more central then $i$. In fact, this metric does not recognise that the path from $l$ to $k$ is not time-respecting. Therefore, the ranking would reflect itineraries that cannot actually be travelled. See Methods for the exact computations. \\
The example in figure \ref{fig:examples}b, instead, shows that without secondary nodes the choice of the length $\Delta \tau$ of a time frame  can affect the ranking, while with Trip Centrality it has no effect, as long as it is shorter than the shortest route duration. In the example, $i$ is connected to $j$ by a link of duration $d$ and $j$ is connected to $k$ by a link of duration $n d$, appearing after an interval $m d$ from the disappearance of the first link. For simplicity, we take $n$ and $m$ integers. The outgoing centrality of node $i$ should clearly be larger than the outgoing centrality on node $j$. However, while Trip Centrality agrees with this intuition, the ranking obtained with the centrality proposed in Ref.[\citen{Grindrod2011}] depends on the choice of $\Delta \tau$, if $\alpha\le(n-1)/2n$. In particular, if $\Delta \tau$ is larger than a certain threshold, $j$ will be ranked as more central than $i$. For example, if $n=2$ and $\alpha\le 1/4$, choosing $\Delta \tau$ shorter than $d/2$ would result in ranking $j$ as the most central. See Methods for details. \\
These two simple examples make clear that, when links have a non-zero link travel time, the introduction of secondary nodes is necessary in order to compute centralities that actually reflect the itineraries that can be travelled on the network. 

\subsection{Application to the Air Transport System}
\label{sec:Data}

\subsubsection{Data}
The dataset used for the following analysis was obtained from the US Department of Transportation's (DOT) Bureau of Transportation Statistics and contains the flights operated in April 2015 by 14 major US airlines. See Methods for further details.\\
The dataset comprises 322 airports. The temporal multiplex is therefore made of 14 layers (one per airline), each with 322 primary nodes. The length of a time frame is chosen as $\Delta \tau=20$ min. 

\subsubsection{Comparison of airports' ranking for different values of $\varepsilon$}
\begin{figure}[t]
\begin{center}
\makebox[\textwidth][c]{\includegraphics{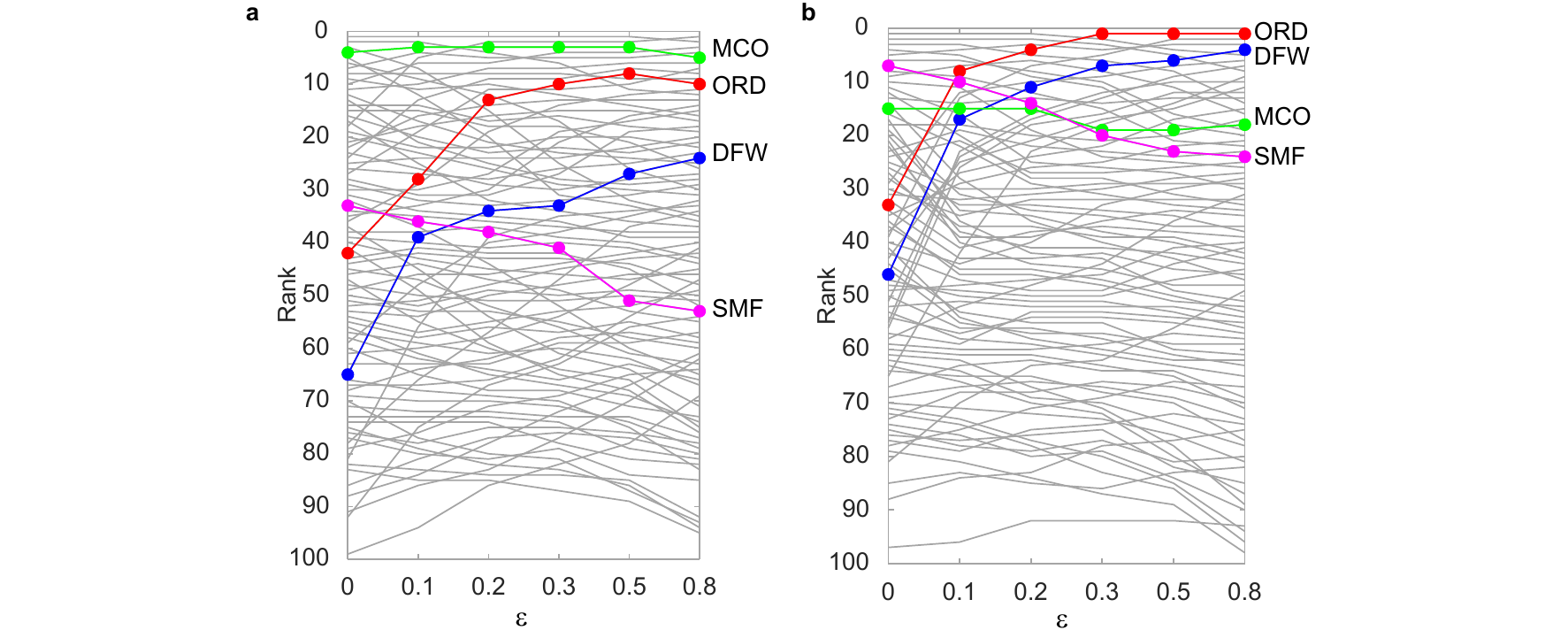}}%
\caption{Evolution of the airports' ranking according to outgoing (a) and incoming (b) Trip Centrality on the scheduled network on April 1st for $\alpha=0.2$ and different values of $\varepsilon$. Only the airports ranking 100 or higher are represented. Each line represents one airport, and the position on the $y$-axis indicates its rank for each $\varepsilon$ value. Rank 1 corresponds to the most central airport. See text for comments on the highlighted lines.}
\label{fig:comparison_e}
\end{center}
\end{figure}

The parameter $\varepsilon \in [0,1]$ measures the propensity of travellers to use different airlines for different legs of one trip. When $\varepsilon$ increases, centralities  always increase, as the weight of inter-layer walks is increased. However, the airports' ranking may change, as some airports rely more on inter-layer walks for their connectivity.  Figure \ref{fig:comparison_e} shows how the airports' ranking according to outgoing and incoming Trip Centrality on April 1st changes when $\varepsilon=0, 0.1, 0.2, 0.3, 0.5, 0.8$, for the airports ranked 100 or higher (for the behaviour of all airports see Supplementary Fig. S1). Results for other days of April are qualitatively similar. Note that Trip Centrality is computed with $\alpha=0.2$, therefore when $\varepsilon<0.2$ a walk using $n$ flights with one inter-layer jump is weighted less than an intra-layer walk made of $n+1$ flights. The contrary is true when $\varepsilon>0.2$. Some airports increase steadily their rank when $\varepsilon$ increases. These are the airports which would gain from an increased cooperation of airlines, making it easier for passengers to use inter-layer walks. Examples of airports gaining rank according to both incoming and outgoing Trip Centrality on every day of April are the Chicago O'Hare Airport (highlighted in red in the figures), the Dallas-Fort Worth International Airport (highlighted in blue), the George Bush Houston Airport, the J. F. Kennedy International Airport. Other airports either maintain their rank, e.g. the Orlando International Airport (green), or decrease it, e.g. the Sacramento International Airport (magenta). Note that such rank decrease is not due to a decrease of the airports' centrality, but simply to their being overtaken by others. \\

\subsubsection{Comparing the network of scheduled and realised flights}
\label{sec:sched_real}
\begin{figure}[t]
\begin{center}
\makebox[\textwidth][c]{\includegraphics{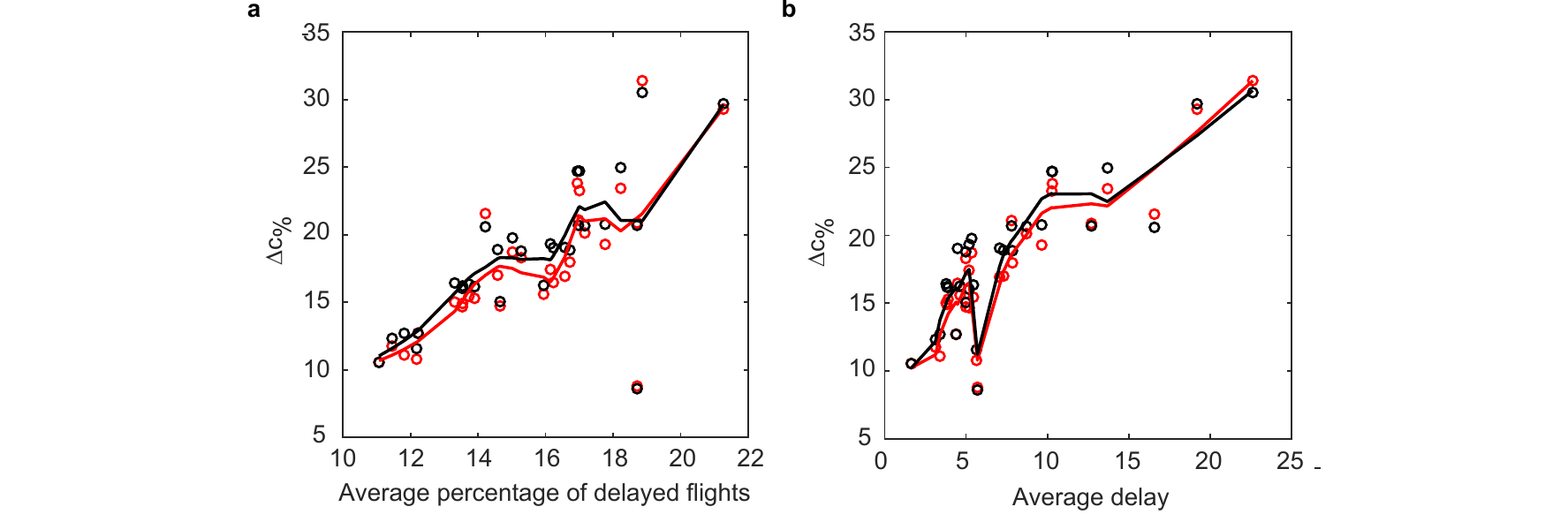}}%
\caption{Percentage of centrality loss, averaged over all airports, in each day of the dataset, according to incoming Trip Centrality (red) and outgoing Trip Centrality (black) plotted against average departure delay (a) and average fraction of flights with departure delay in one airport (b). Trip centrality is computed with $\alpha=0.2$ and $\varepsilon=0$. Each point corresponds to one day of the dataset. The percentage centrality loss of an airport is computed as $\Delta c_{\%}= 100 \times (c_{sched}-c_{act})/c_{sched}$, where $c_{sched}$ and $c_{act}$ are the airport's centralities on the scheduled and realised network. Lines are obtained by a locally weighted smoothing (LOWESS) of the dots of the correspondent colour.}
\label{fig:av_cent_loss}
\end{center}
\end{figure}

\begin{figure}[t]
\begin{center}
\makebox[\textwidth][c]{\includegraphics{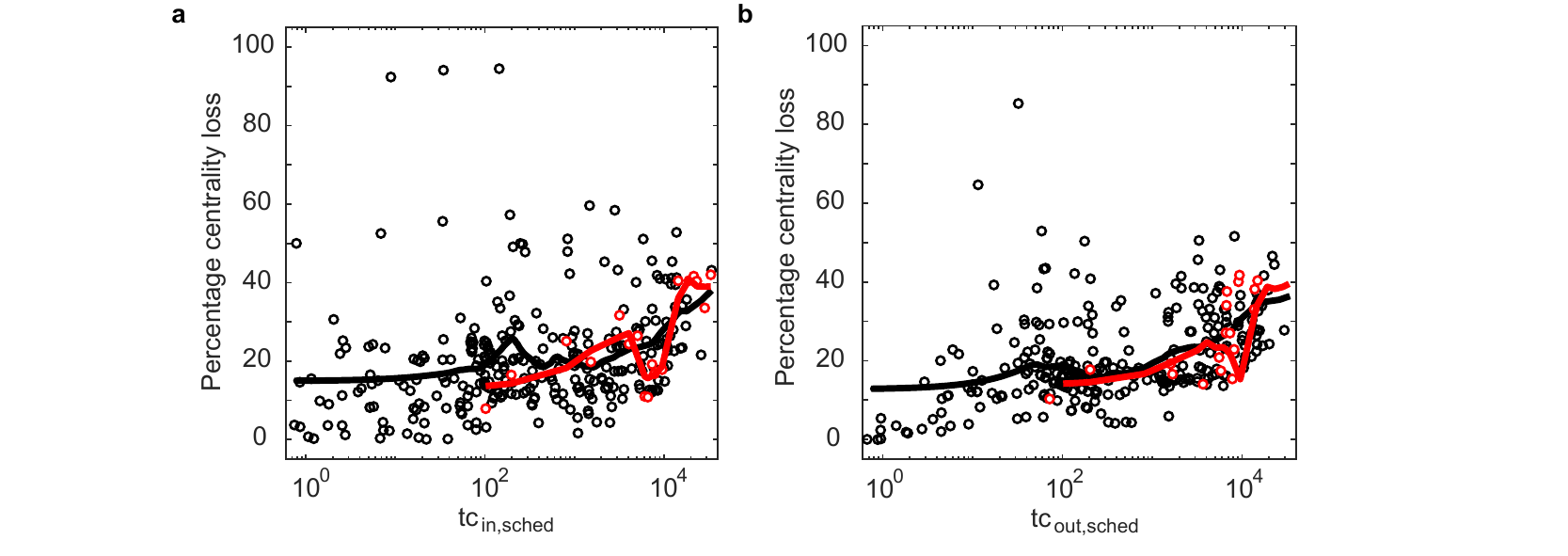}}%
\caption{Percentage of Trip Centrality loss plotted against Trip Centrality in the scheduled network, for April 1st. a) incoming, b) outgoing. Trip centrality is computed with $\alpha=0.2$ and $\varepsilon=0$. Red dots represent distressed airports (see text for definition), black dots non-distressed ones. Lines are obtained by a locally weighted smoothing (LOWESS) of the dots of the correspondent colour. }
\label{fig:distressedVSnon}
\end{center}
\end{figure}

In many situations, especially in transportation, there exists a scheduled and a realised network, differing due to delays and cancellations. Delays can cause the disruption of connections between flights which were feasible in the scheduled network, therefore diminishing the potential to move through the network. In this section, we show that Trip Centrality is able to quantify the loss of connectivity of a node due to the delays, differently from static centrality metrics, which do not account for the temporal structure. \\
For each day, we compute each airport's aggregated Trip Centrality on the scheduled and realised networks. Note that delays can also cause the opening of new walks on the realised network, which were not present in the scheduled one. While in bus or metro networks these new walks could be used by passengers, in air traffic this is not possible, therefore these `forbidden' walks should be excluded from the computation. The procedure to do so is detailed in the Methods section.
Then, the loss of centrality of airport $i$ is computed as the difference between its centrality in the scheduled and realised network. Figure \ref{fig:av_cent_loss} plots the percentage centrality loss, averaged over all airports, for each day against two delay-related indicators characterising that day: the average delay of all flights on that day and the average fraction of delayed flights in an airport. The average percentage centrality loss increases with both indicators, meaning that, in aggregate, more or larger delays in the network cause larger centrality losses. Instead, the correlation between the rankings on the scheduled and realised networks decreases with both indicators (see Supplementary Fig. S2). These effects are due both to delays and to cancelled and diverted flights (which are increasing with both indicators). However, Supplementary Fig. S3 shows that the patterns are still present when cancelled and diverted flights are excluded from the analysis, proving that the proposed centrality metric is able to detect the changes in the network's connectivity due to delays, or, more in general, to a change in the links' temporal structure. Results are robust with respect to changes of $\alpha$ (see Supplementary Figs. S4 and S5). \\
While average centrality losses tend to be larger in days with more or larger delays, at the level of a single airport they are weakly correlated with delays in that airport. Let us call an airport \textit{distressed} when the fraction of its flights with delay larger than average is larger than the average fraction of delayed flight in an airport on the analysed day. Figure \ref{fig:distressedVSnon} shows that the percentage centrality losses of distressed airports are not larger than those of non-distressed ones. The figures refer to April 1st, results for other days are similar. Such results underline that the loss of Trip Centrality reflects the network effects of delay, which do not depend simply on the delay itself. In fact, the same delay can have a large network effect if it causes the disruption of several connections, or no effect at all if no connection is disrupted. While larger delays have a larger probability to cause a disruption, there is no simple relationship between the amount of delay and the number of missed connections, and therefore centrality loss, which depend on the schedule. 

\subsubsection{Comparison of existing centrality metrics with Trip Centrality}
\begin{figure}[h!]
\begin{center}
\makebox[\textwidth][c]{\includegraphics{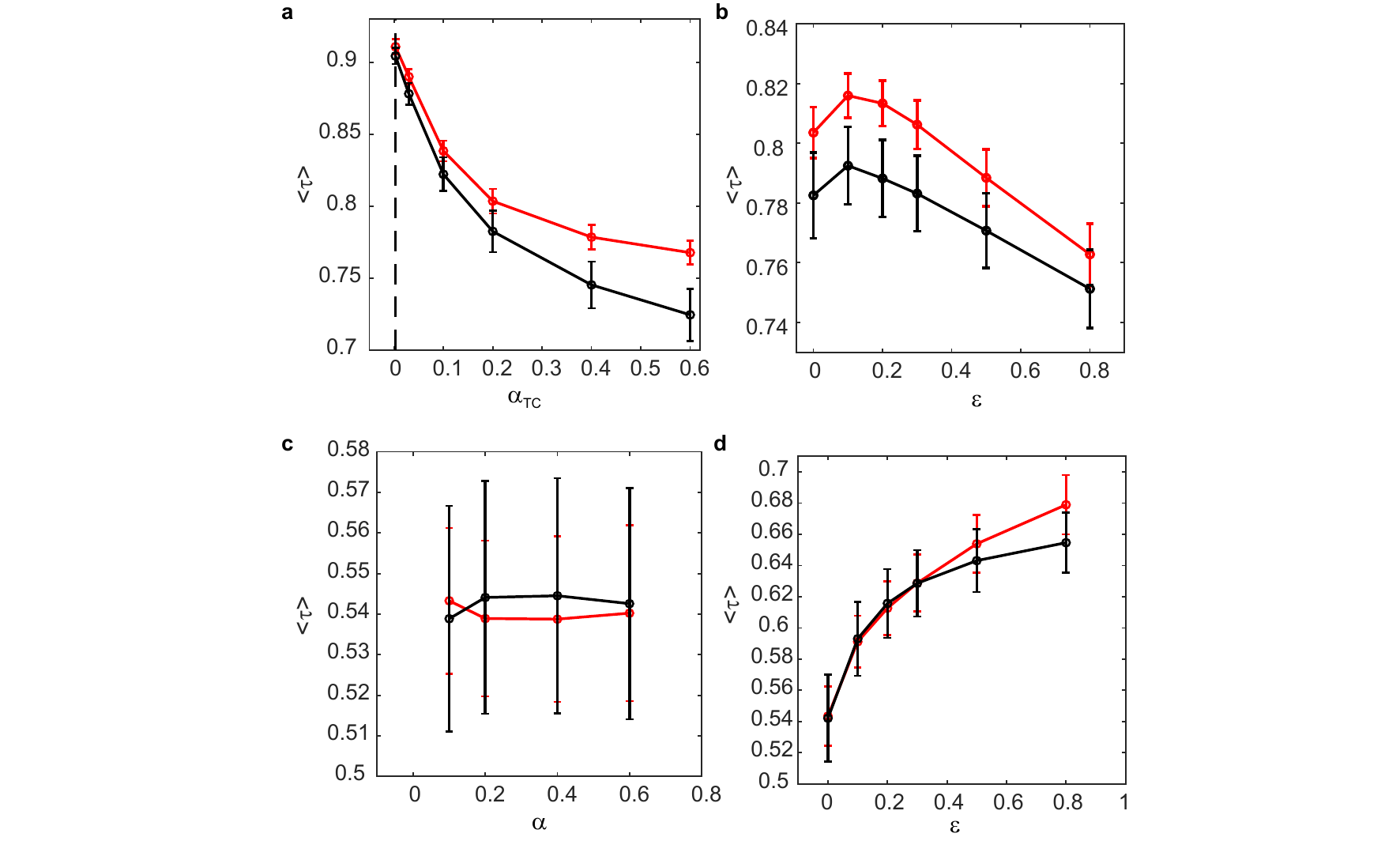}}%
\caption{a) Kendall correlation coefficient, averaged on all days, between the rankings according to Katz centrality with $\alpha=0.003$ and Trip Centrality, for different values of $\alpha$ used in the computation of Trip Centrality. The red line corresponds to the incoming centralities and the black one to the outgoing. The dotted line marks the value of $\alpha$ used for Katz centrality;
b) Kendall correlation coefficient, averaged on all days, between the rankings according to Katz centrality with $\alpha=0.003$ and Trip Centrality with $\alpha=0.2$, for different values of $\varepsilon$ used in the computation of Trip Centrality. Colors as in panel a;
c) Kendall correlation coefficient, averaged on all days, between the rankings according to Dynamic communicability and Trip Centrality for different values of $\alpha$. Trip Centrality is computed with $\varepsilon=0$. Colors as in panel a;
d) Kendall correlation coefficient, averaged on all days, between the rankings according to Dynamic communicability and Trip Centrality with $\alpha=0.2$, for different values of $\varepsilon$ used in the computation of Trip Centrality. Colors as in panel a. Bars represent standard errors. }
\label{fig:statVStrip}
\end{center}
\end{figure}

Here, we compare the ranking of airports according Trip Centrality with those according to Katz centrality and Dynamic communicability, in order to show their difference. Rankings are compared computing the Kendall rank correlation coefficient $\tau$, which measures the similarity of two ranked sequences. The coefficient takes values in [-1,1], with 1 corresponding to two identical sequences and -1 to two sequences that are one the inverse of the other.\\
For Katz centrality we choose the largest value of $\alpha$ ensuring convergence, $\alpha=0.003$. We remark that such very small value strongly penalises long walks (more than 300 walks of length 2 are needed to contribute as much as one walk of length 1), making the ranking according to Katz centrality quite similar to a simple ranking according to the number of flights (average correlation coefficient 0.9, see Supplementary Fig. S6 for a comparison of the rankings for April 1st). Dynamic communicability is computed according to equation \eqref{eq:grindrog}, where no restriction on $\alpha$ applies. 
All metrics are computed on the scheduled network. \\
Katz centrality and Trip Centrality produce similar rankings, on average, only when the latter is computed with a very small $\alpha$ (see Fig. \ref{fig:statVStrip}a), as walks longer than one (in whose counting they differ) have a negligible weight. When Trip Centrality is computed with $\alpha=0.003$, i.e. the same as Katz centrality, and $\varepsilon=0$, the average correlation coefficient is $0.91 \pm 0.01$, both for the incoming and the outgoing case. However, they become less and less similar when the value of $\alpha$ used in the computation of Trip Centrality becomes larger, increasing the importance of longer walks. Increasing the value of $\varepsilon$ above 0.1 decreases the similarity (see Fig. \ref{fig:statVStrip}b). Supplementary Fig. S7a shows, as an example, the comparison of the rankings according to incoming Katz and Trip Centrality for April 1st, with Trip Centrality's parameters $\alpha=0.2$ and $\varepsilon=0$. \\
Dynamic communicability and Trip Centrality, computed with the same value of $\alpha$, become more similar when $\varepsilon$ approaches 1 (see Fig. \ref{fig:statVStrip}d, with $\alpha=0.2$). This is understandable, since Dynamic communicability does not distinguish layers. However, there does not exist a value of $\varepsilon$ for which they are very similar. For $\alpha=0.2$, for example, they approach an average correlation of $\tau=0.7$ when $\varepsilon$ approaches 1. The value of $\alpha$ does not influence much their correlation (see Fig. \ref{fig:statVStrip}c). Supplementary Figure S7b shows, as an example, the comparison of the rankings according to incoming centrality for April 1st, with $\alpha=0.2$ and Trip Centrality with $\varepsilon=0$. Interestingly, for any parameter choice, Trip Centrality is more similar to Katz than to Dynamic communicability. This is explained by the fact that Trip and Katz centralities agree on the counting of one-flight itineraries, which give the larger contributions, while with Dynamic communicability longer flights count more than shorter ones, as they appear in more time frames.   \\
To summarise, when itineraries of more than one flight are assigned a non-negligible weight, i.e. $\alpha$ not too small, accounting for the temporal multiplex structure of the network and for the non-zero link travel time does indeed make a significant difference in the ranking. \\
The ten most central airports according to each metric are reported in Supplementary table 1.\\

\section{Discussion}
\label{sec:disc}
The new centrality metric we introduced generalises centrality to networks where links are dynamic, characterised by a non-zero travel time and of different types. With respect to existing centrality metrics, Trip Centrality assigns centrality to nodes according to the schedule-respecting itineraries that can be travelled on the network, with the possibility of weighting less the inter-layer ones, which use links of several types. The importance of accounting for the links' duration, through the introduction of secondary nodes, was proved by two simple examples. The metrics retain the computational convenience of the original static metrics, requiring only simple matrix multiplications. Additionally, with respect their static counterparts, there is no constraint on the value of the weight $\alpha$ to assign to a link, as there is no convergence issue. Therefore, $\alpha$ can be freely chosen as the value better reflecting the relative importance of walks of different length. \\
Transportation networks are the most direct application for Trip Centrality, which is however more general, being suited for any temporal network with non-zero link travel time. For example, our approach could be applied to study epidemic spreading on a temporal network when an incubation time is required before an infected individual can infect susceptible ones, or the diffusion of information by email, when emails are not read instantaneously. \\
Moreover, the proposed approach to combine the multiplex structure with the temporal one, i.e. the use of the matrix $K$ to weight a change of layer, can be used even in the case of instantaneous transportation by applying it without the introduction of secondary nodes, i.e. by using adjacency matrices $A^{[t]}$ of size $N N_L \times N N_L$ describing links among primary nodes, distinguished by layers. We remark that Trip Centrality, if applied on a single layer and without secondary nodes, reduces to the version of Dynamic communicability allowing only one jump per time frame. \\
Trip Centrality can be used to assess the change in a node's centrality following the change of the links' time dynamics, and also to explore how the interaction of different layers (parameter $\varepsilon$) affects the centralities. The application to the US air transport system showed that a subset of the airports, typically large hubs, would be particularly favoured by an improved cooperation among airlines, promoting the use of itineraries comprising flights of different airlines. Additionally, Trip metrics proved able to capture that, on average, the connectivity of the airports and flights network diminishes the more delays there are on the network (i.e. the average percentage centrality loss becomes larger). This could not be captured by static metrics, neglecting the time dynamics. However, we showed that, for single airports, the percentage centrality loss is weakly correlated with delays in the airport (since the behaviour of distressed and non-distressed airports is comparable). This is in agreement with the fact that Trip Centrality loss measures the effect of a delay in terms of loss of network connectivity, which depends in a complex, non-linear way on the delay itself. In fact, depending on the schedule, a smaller delay could create more disruption than a larger one, and an airport might lose incoming centrality even if its incoming flights are on time, due to delays in the previous legs of the incoming itineraries. In conclusion, Trip Centrality provides a tool to assess the effect of delays on the connectivity of transportation networks at the whole network level as well as at the single node level. \\

\section{Methods}
\subsection{Computation of centralities in the example in Fig.\ref{fig:examples}a}
According to Dynamic communicability, the vector of outgoing centralities is 
\begin{equation}
\label{eq:grindrog}
    \vec{c}^{out}=[(\mathbb{I}+\alpha A^{[1]})(\mathbb{I}+\alpha A^{[2]})\dots(\mathbb{I}+\alpha A^{[T]})-\mathbb{I}]\vec{1}_N,
\end{equation}
where $A^{[t]}$ is the $N\times N$ adjacency matrix without secondary nodes. The links present at each time frame in this adjacency matrix are represented in Fig.\ref{fig:examples}a.II. To compute the outgoing centralities of $i$ and $l$ we can count the outgoing walks: $i$ has two outgoing walks of length 1 to $j$, counted by $A^{[1]}$ and $A^{[2]}$, and four of length 2 to $k$, counted by the products $A^{[1]}A^{[3]}$,  $A^{[1]}A^{[4]}$,  $A^{[2]}A^{[3]}$ and $A^{[2]}A^{[4]}$. Its centrality is therefore $c^{out}_i=2\alpha+4\alpha^2$. Instead, $l$ has three outgoing walks of length 1, and still 4 of length 2, therefore its centrality is  $c^{out}_l=3\alpha+4\alpha^2$, larger than the centrality of $i$. \\
Introducing secondary nodes, Trip Centrality can be computed according to equation \eqref{eq:trip_S}, as we are working on a single layer. The adjacency matrices are now $(N+4)\times (N+4)$. The centrality of the two nodes are then, $tc^{out}_i=\tilde{\alpha}+\tilde{\alpha}^2+\tilde{\alpha}^3+\tilde{\alpha}^4=\tilde{\alpha}+\tilde{\alpha}\alpha +\alpha +\alpha^2$ and $tc^{out}_l=\tilde{\alpha}+\tilde{\alpha}^2=\tilde{\alpha} +\alpha$, where we used $\alpha=\tilde{\alpha}^2$. Therefore, Trip Centrality correctly finds that $i$ has a larger outgoing centrality than $l$. \\

\subsection{Computation of centralities in the example in Fig.\ref{fig:examples}b}
Let $\Delta \tau$ be the length of a time frame, such that $\Delta \tau<d$. For simplicity, let us assume that $\Delta \tau=d/\Delta$, with $\Delta$ an integer larger than 1. Therefore, in the formulation without secondary nodes, the link from $i$ to $j$ is present in $\Delta$ time frames and the link from $j$ to $k$ in $n\Delta$. It is then possible to go from $i$ to $j$ with $\Delta$ different walks of length 1, from $j$ to $k$ with $n\Delta$, and from $i$ to $k$ with $n\Delta^2$ different walks of length 2. Therefore we have $c^{out}_i= \alpha \Delta+\alpha^2 n\Delta^2$ and $c^{out}_j=\alpha n\Delta$. Now, $c^{out}_i>c^{out}_j$ when $\Delta>(n-1)/\alpha n$.  If $\alpha>(n-1)/2n$ the condition is always realised (because $\Delta\ge 2$), and $i$ is more central than $j$ for any $\Delta$. If $\alpha \le(n-1)/2n$, instead, depending on the value of $\Delta$, i.e. on $\Delta \tau$, $i$ can be less or as central as $j$. This dependency on $\Delta \tau$ is not present in Trip Centrality, according to which the two centralities are $tc^{out}_i=\tilde{\alpha}+\tilde{\alpha}^2+\tilde{\alpha}^3+\tilde{\alpha}^4$ and $tc^{out}_j=\tilde{\alpha}+\tilde{\alpha}^2$, which do not depend on $\Delta \tau$.

\subsection{US Air Traffic dataset}
The dataset made available by the US Department of Transportation's (DOT) Bureau of Transportation Statistics was obtained from  \url{https://www.kaggle.com/usdot/flight-delays}. It includes all flights operated in 2015 by 14 major US airlines. In the multiplex of airports and flights, each airline is treated as a separate layer. Note that, in general, if two airlines belong to the same alliance they should be included in the same layer. In fact, from the passengers' point of view, connections between their flights are analogous to within-airline connections. None of the 14 airlines of the US dataset, however, belong to the same alliance. \\
The present analysis was performed on the month of April, which is quite heterogeneous in terms of delays, ranging from days with few small delays to days with many large ones. For each flight in the dataset, the following information is available: Date, Flight number, Tail number, Origin airport, Destination airport, Scheduled departure time, Realised departure time, Scheduled arrival time, Realised arrival time, Airline, Cancellation (1 if was cancelled, 0 otherwise), Diverted (1 if it was diverted, 0 otherwise). All times are converted from local time to Eastern Standard Time (EST). Days are considered to start at 4AM EST, as this is the time of minimum traffic across the entire US.

\subsection{Loops}
The proposed generalization of Katz centrality, as its static counterpart, counts walks on the network. Therefore, it includes walks that go back to a node more than once. Depending on the application, this might be wanted or not. In the air transport case, walks with loops might be important for delay propagation, but certainly not for passengers. However, it can be shown that eliminating all walks that  pass a second time from their starting point leaves the ranking almost unchanged. Therefore, we claim that considering or not loops does not change significantly the results obtained. \\
The elimination of walks that pass a second time from their starting point is obtained as follows. Let
\begin{equation}
    Q=(\mathbb{I}+\tilde{\alpha}A^{[1]}K)\dots(\mathbb{I}+\tilde{\alpha}A^{[T]}K).
\end{equation}
With this definition,
\begin{subequations}
\begin{align}
 \vec{t}^{out}&=[Q-\mathbb{I}]K^{-1}\vec{1}_{N N_L+N_l}, \\ 
 \vec{t}^{in}&=\vec{1}_{N N_L+N_l}^T[Q-\mathbb{I}]K^{-1}.
\end{align}
\end{subequations}
The element $Q_{ik}$ contains the contribution to the outgoing centrality of $i$ given by walks from $i$ to $j$ (or equivalently the contribution to the incoming centrality of $j$ given by walks from $i$ to $j$). If we compute the matrix $Q$ one time frame at a time, $Q^{[1]}=(\mathbb{I}+\tilde{\alpha} A^{[1]}K)$ only counts the walks during the first time frame, $Q^{[2]}=(\mathbb{I}+\tilde{\alpha}A^{[1]}K)(\mathbb{I}+\tilde{\alpha}A^{[2]}K)$ counts the walks up to the second time frame, and so on. At each time step, the diagonal elements have value $Q^{[t]}_{ii}=1+\mathrm{contribution \ of\  walks \ going \ from\  }i\mathrm{ \ to\  }i.$ By setting the diagonal elements to 1 after each multiplication, we eliminate the contributions of walks coming back to their starting point during that time frame. This eliminates also all the walks which would be the continuation of such loops. 

\subsection{Centrality in the realised network}
\label{sec:realised}
We call $\{A^{[t]}_{sched}\}_{t=1,...,T}$ and $\{A^{[t]}_{real}\}_{t=1,...,T}$, respectively, the adjacency matrices defined at each time frame for the networks of scheduled and realised flights. In the realised network the timing of links is changed due to delays, cancellations and diversions, therefore some of the time-oriented walks that existed in the scheduled network will not be present anymore, causing losses of centrality with respect to the scheduled network. Note that, in principle, delays could also create new walks, by allowing connections that were impossible in the scheduled network. In the case of air transport, differently from the bus or metro networks, such new walks cannot be used by passengers (except in the case of rerouting). Therefore, if we want to assess the connectivity of an airport from the passengers' point of view, we should not consider the contribution of these `forbidden' walks. The forbidden walks can appear in two cases. The first case is when a negative arrival delay (early arrival) allows a connection with a departing flight. In this case, the problem is solved by setting the negative arrival delay to zero, i.e., setting the arrival time equal to the scheduled arrival time. The second case is when the delay of a departing flight allows the connection with an incoming flight, which was originally landing too late to make the connection. A partial solution which eliminates most of the forbidden walks can be implemented and is explained in section 4 of the SI. 

\section{Acknowledgments}
This project has received funding from the SESAR Joint Undertaking under the European Union’s Horizon 2020 research and innovation programme under grant agreement No 783206. The opinions expressed herein reflect the authors’ views only. Under no circumstances shall the SESAR Joint Undertaking be responsible for any use that may be made of the information contained herein.

\section{Author contributions}
SZ, PM and FL designed the research. SZ performed the research and the data analysis, with assistance from PM and FL. All authors contributed to the writing and to the discussion of the results.

\section{Competing interests statement}
The authors declare no conflict of interest.

\newpage

\bibliographystyle{naturemag}
\bibliography{TripCentrality.bib}

\end{document}


\maketitle

\thispagestyle{plain}

\section{Computing Trip Centrality when $\varepsilon=1$}
When $\varepsilon=1$, the matrix $K$ used in the computation of Trip Centrality (equation (4) of the main text) is not invertible, therefore Trip Centrality cannot be computed using equation (4). In this section, we prove that, if the final multiplication by $K^{-1}$ in equation (4) is skipped, the rankings according to the single-layer outgoing centrality and the aggregated outgoing and incoming centralities do not change (although the values of the centralities do change). We conclude, therefore, that the rankings according to these centralities for the case $\varepsilon=1$, computed skipping such final multiplication, can be safely compared with the rankings obtained with equation (4) for other values of $\varepsilon$. \\
Consider Trip Centrality, and let
\begin{equation}
    M=[(\mathbb{I}+\tilde{\alpha}A^{[1]}K)\dots(\mathbb{I}+\tilde{\alpha}A^{[T]}K)- \mathbb{I}] K^{-1}.
\end{equation}
 
The element $M_{ij}$ contains the sum of the weights of walks of all lengths from $i$ to $j$. The incoming Trip Centrality of the node $j$ is then obtained as $t^{IN}_j=\sum_i M_{ij}$ and the outgoing Trip Centrality of the node $i$ as $t^{out}_i=\sum_j M_{ij}$. \\
Let $\tilde{M}$ be the equivalent of $M$ but skipping the final division by $K$,
\begin{equation}
    \tilde{M}=[(\mathbb{I}+\tilde{\alpha}A^{[1]}K)\dots(\mathbb{I}+\tilde{\alpha}A^{[T]}K)- \mathbb{I}].
\end{equation}
 
The element $\tilde{M}_{ij}$ counts, on top of the walks counted by $M_{ij}$, also the walks that arrived to one of the copies of $j$ on other layers and then jumped to $j$ in the last step, with an additional weight $\varepsilon$ due to the final change of layer.  Calling $C_j$ the set of copies of $j$, we have thus
\begin{equation} 
\tilde{M}_{ij}=M_{ij}+\varepsilon \sum_{k\in C_j \setminus \{j\}}M_{ik}=(1-\varepsilon) M_{ij}+\varepsilon \sum_{k\in C_j }M_{ik} .
\end{equation}
Note that, if $j$ is a secondary node, $C_j=\{j\}$, and $\tilde{M}_{ij}=M_{ij}$. The single-layer centralities computed from $\tilde{M}$ are, therefore:
\begin{equation}
\begin{split}
\tilde{t}^{in}_j&=\sum_i \tilde{M}_{ij}= (1-\varepsilon) t^{in}_j+ \varepsilon \sum_i \sum_{k\in C_j }M_{ik}= (1-\varepsilon) t^{in}_j+ \varepsilon  \sum_{k\in C_j }t^{in}_k,
\end{split}
\end{equation}
\begin{equation}
\begin{split}
\tilde{t}^{out}_i&=\sum_j \tilde{M}_{ij}= (1-\varepsilon) t^{out}_i+ \varepsilon \sum_j \sum_{k\in C_j }M_{ik}= (1-\varepsilon) t^{out}_i +\varepsilon N_L t^{out}_i \\
&=(1+ (N_{L} -1)\varepsilon) t^{out}_i,
\end{split}
\end{equation}
where $N_L$ is the number of layers. \\
In the outgoing case, $t$ and $\tilde t$ differ only by a multiplicative constant, therefore multipliying or not by $K^{-1}$ produces the same ranking. This is not true, however, for the incoming case (for single-layer centrality). \\
Let us now check what happens for the aggregated centralities of primary nodes.
\begin{equation}
\tilde{t}^{in,a}_j= \sum_{k\in C_j} \tilde{t}^{in}_k= \sum_{k\in C_j}[ (1-\varepsilon) t^{in}_k+ \varepsilon  \sum_{l\in C_k }t^{in}_l] = (1-\varepsilon) t^{in,a}_j +\varepsilon N_{L}  t^{in,a}_j = (1+ (N_{L} -1)\varepsilon)t^{in,a}_j,
\end{equation}
\begin{equation}
\tilde{t}^{out,a}_i= \sum_{k\in C_i} \tilde{t}^{out}_k= \sum_{k\in C_i} (1+ (N_{L} -1)\varepsilon) t^{out}_k = (1+ (N_{L} -1)\varepsilon) t^{out,a}_i.
\end{equation}
We obtain that, for both the incoming and the outgoing case, the aggregated centrality with or without multiplication by $K^{-1}$ differ only by a multiplicative constant, therefore produce the same ranking.

 \section{TripRank: a generalisation of PageRank to temporal multiplexes}
 PageRank is a generalisation of Katz centrality, developed by Google \cite{Brin1998}, that introduces an additional weight to the walks, depending on the in- (or out-) degree of the nodes they cross. Considering a static directed network of $N$ nodes with weighted adjacency matrix $A_{ij}$, such that $A_{ij}=k$ if there are $k$ links from $i$ to $j$. The incoming PageRank centrality of node $i$ is given by the sum of the contributions of walks of any length incoming to $i$, where each walk of length $n$ contributes $\alpha^n$ times a factor $1/d^{out}_l$ for each node $l$ crossed by the walk, with $d^{out}_l$ the out-degree of node $l$. This sum can be computed as  
\begin{equation}
\label{eq:pagerank}
 pr_i^{in}=\sum_{n=0}^\infty \sum_{j=1}^N (\alpha D^{-1} A)^n_{ji}  =\sum_{j=1}^N(\mathbb{I}-\alpha D^{-1}A)^{-1}_{ji},  
\end{equation}
where $D_{ij}=\delta_{ij}d_i^{out}$ with $\delta_{ij}$ the Kronecker delta, so that using a link from $j$ to $k$ is weighted by the inverse of the out-degree of $j$, $1/d_j^{out}$. Convergence requires $\alpha$ to be smaller than the maximum eigenvalue of $D^{-1}A$. In terms of our air traffic example, the meaning of this generalization is that an airport with an inbound flight coming from a large airport, with a large out-degree, will inherit a fraction of its centrality proportional to the inverse of such out-degree. In other words, the more outbound flights an airport has, the less of its centrality such destination airport inherits. In the out going case, similarly we have  
 \begin{equation}
 pr_i^{out}=\sum_{j=1}^N(\mathbb{I}-\alpha A D^{-1})^{-1}_{ji},  
\end{equation}
where $D_{ij}=\delta_{ij}d_i^{in}$. Convergence requires $\alpha$ to be smaller than the maximum eigenvalue of $A D^{-1}$.\\

The single-layer temporal generalisation of PageRank, here termed \textit{Single-Layer TripRank},  is obtained analogously to Single-Layer Trip Centrality as
\begin{subequations}
\begin{align}
    \vec{tr}_S^{out}&=[(\mathbb{I}+\tilde{\alpha} A^{[1]}D_{in}^{-1})(\mathbb{I}+\tilde{\alpha}  A^{[2]}D_{in}^{-1})\dots(\mathbb{I}+\tilde{\alpha}  A^{[T]}D_{in}^{-1})-\mathbb{I}]\vec{1}_{N+N_l},\\
    \vec{tr}_S^{in}&=\vec{1}_{N+N_l}^T[(\mathbb{I}+\tilde{\alpha} D_{out}^{-1} A^{[1]})(\mathbb{I}+\tilde{\alpha}  D_{out}^{-1} A^{[2]})\dots(\mathbb{I}+\tilde{\alpha}  D_{out}^{-1} A^{[T]})-\mathbb{I}],
\end{align}
\end{subequations}
where the temporal adjacency matrices $A^{[t]}$ are defined as in the main text and $D_{in}$ and $D_{out}$ are the diagonal matrices $D_{in,ij}=\delta_{ij}d^{in}_i$ and $D_{out,ij}=\delta_{ij}d^{out}_i$, with $d^{in}_i$ and $d^{out}_i$, respectively, the number of incoming and outgoing links of node $i$ during the entire observation window.\\
The temporal multiplex generalisation, named TripRank, is also obtained analogously to Trip Centrality:
\begin{subequations}
\begin{align}
\vec{tr}^{out}&=[(\mathbb{I}+\tilde{\alpha} A^{[1]}D_{in}^{-1}K)(\mathbb{I}+\tilde{\alpha}  A^{[2]}D_{in}^{-1}K)\dots(\mathbb{I}+\tilde{\alpha}  A^{[T]}D_{in}^{-1}K)-\mathbb{I}]K^{-1}\vec{1}_{N N_L+N_l},\\
 \vec{tr}_S^{in}&=\vec{1}_{N N_L+N_l}^T[(\mathbb{I}+\tilde{\alpha} D_{out}^{-1} A^{[1]}K)(\mathbb{I}+\tilde{\alpha}  D_{out}^{-1} A^{[2]}K)\dots(\mathbb{I}+\tilde{\alpha}  D_{out}^{-1} A^{[T]}K)-\mathbb{I}],
 \end{align}
\end{subequations}
where the matrix $K$ and the multilayer adjacency matrices $A^{[t]}$ are defined as in the main text and $D_{in}$ and $D_{out}$ are the diagonal matrices $D_{in,ij}=\delta_{ij} \tilde{d}^{in}_{i}$ and $D_{out,ij}=\delta_{ij}\tilde{d}^{out}_{i}$, with $\tilde{d}^{in}_{i}$ and $\tilde{d}^{out}_{i}$, respectively, the number of incoming and outgoing links of node $i$ or any of its copies on the other layers during the entire observation window.\\

\section{Results of US air traffic datasets analysis with TripRank}

\subsection{Comparison between Trip Centrality and TripRank}
\begin{figure}[h!]
\begin{center}
\makebox[\textwidth][c]{\includegraphics{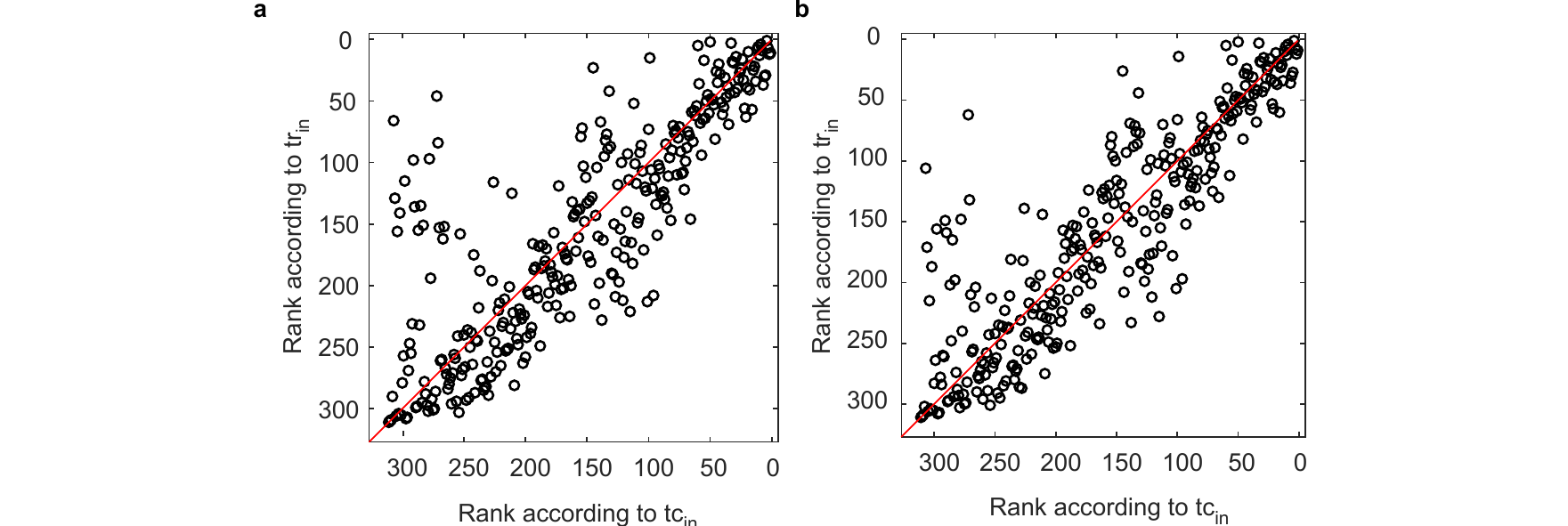}}%
\caption{Comparison of the rankings according to incoming Trip Centrality and to TripRank for April 1st, on the scheduled network and with $\varepsilon=0$. In panel a), both metrics are computed with $\alpha=0.2$, in panel b) TripRank is computed with $\alpha=0.5$. The red line is the 1:1 line.}
\label{fig:tcVStr}
\end{center}
\end{figure}

The rankings of airports according to Trip Centrality and TripRank are quite different: with $\varepsilon=0$ and $\alpha=0.2$, they have a Kendall correlation coefficient, averaged over all days in the dataset, $\tau=0.70 \pm 0.01$ in the incoming case and $\tau=0.69 \pm 0.02$ in the outgoing case. Increasing $\varepsilon$ the correlation decreases further. TripRank penalises long walks more than Trip Centrality does, as each additional flight brings a weight $\alpha/d$, with $d$ the degree of its departure or arrival airport (depending on whether we consider incoming or outgoing TripRank), typically much smaller than $\alpha$. Therefore, it is not surprising that increasing $\varepsilon$, which means that more long walks are considered and increasingly weighted, decreases the similarity of the two metrics. For the same reason, the correlation coefficient of the two metrics increases if TripRank is computed with a larger $\alpha$, for example for $\varepsilon=0$ and $\alpha=0.8$ for TripRank, we find $\tau=0.76\pm 0.01$ in the incoming case and $\tau=0.74\pm 0.01$ in the outgoing one. Supplementary Fig. \ref{fig:tcVStr}a shows the comparison of the rankings for April 1st, with $\varepsilon=0$ and $\alpha=0.2$, in the incoming case. It can be seen that the larger differences are given by airports gaining a large number of ranks with TripRank. Similar results are observed for the other days in the dataset. The airports gaining many ranks with TripRank are typically the ones that are connected to airports with a very small degree, e.g. the airport of Anchorage, with flights to other small regional airports in Alaska. In fact, these kind of connections in the case of TripRank assure a relatively high centrality contribution, while according to Trip Centrality they give the same contribution as a connection to a large airport, which however brings additional centrality through longer walks, favouring airports connected to large airports. Increasing the value of $\alpha$ used to compute TripRank reduces these differences (see supplementary Fig. \ref{fig:tcVStr}b), however the rankings remain fundamentally different. 

\subsubsection{Comparison of TripRank and PageRank}
In this section, we compare the ranking of airports according to TripRank and to its static single-layer counterpart, PageRank. PageRank is computed with the largest value of $\alpha$ allowing convergence, $\alpha=0.19$. The Kendall correlation coefficient between the rankings according to the two metrics computed using the same value of $\alpha$ is $\tau=0.80\pm 0.02$ in the incoming case and  $\tau=0.79\pm 0.02$ in the outgoing case when $\varepsilon=0$, and grows with increasing $\varepsilon$, as expected given that PageRank does not distinguish layers (see supplementary Fig. \ref{fig:prVStr_corr}). A comparison of the rankings for April 1st, for the incoming case, is shown in supplementary Fig. \ref{fig:prVStr}.  For $\varepsilon$ close to 1, the two ranking become quite similar (Fig.  \ref{fig:prVStr_corr}b). Results are qualitatively similar for other days of the dataset. The correlations increase very slightly for increasing values of $\alpha$ used in the computation of TripRank. The top ten airports according to the two centrality metrics are reported in the supplementary table \ref{tab:topten_TR}.

\subsection{Comparison of airports' ranking for different values of $\varepsilon$}
\begin{figure}[h!]
\begin{center}
\includegraphics{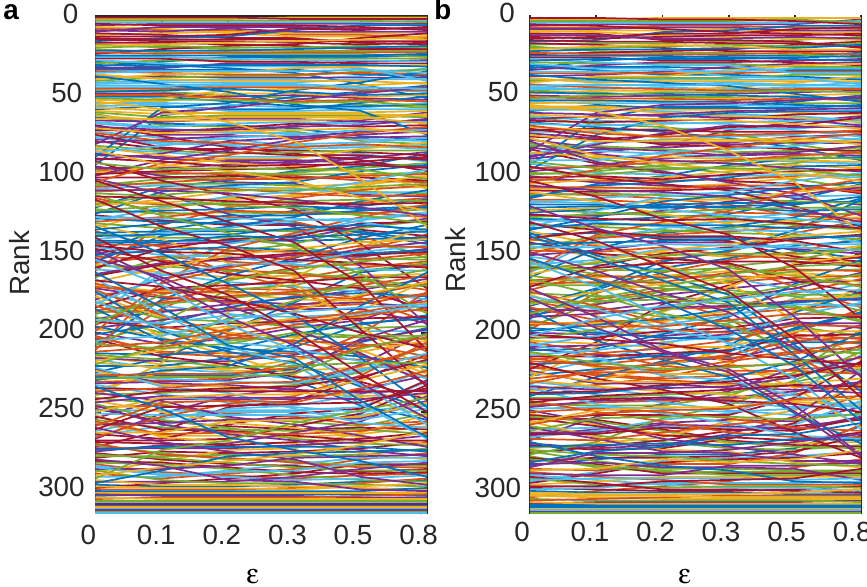}
\caption{Evolution of the airports' ranking according to outgoing (a) and incoming (b) TripRank on the scheduled network on April 1st for different values of $\varepsilon$. TripRank is computed with $\alpha=0.5$. Each line represents one airport, and the position on the $y$-axis indicates its rank for each $\varepsilon$ value. Rank 1 corresponds to the most central airport. }
\label{fig:suppl_comparison_e}
\end{center}
\end{figure}

Figure \ref{fig:suppl_comparison_e} shows how the airports' ranking according to outgoing and incoming TripRank on the scheduled network on April 1st changes when $\varepsilon=0, 0.1, 0.2, 0.3, 0.5, 0.8$.  Results for other days of April are qualitatively similar. Differently from what was shown in section 2.3.2 of the main text for Trip Centrality, in this case the most central airports are quite stable when $\varepsilon$ varies, and in general rank increases are smaller than with Trip Centrality. This is explained by the fact that the percentage variations of centrality values caused by changes of $\varepsilon$ are much smaller than those found with Trip Centrality, as long walks are more penalised by TripRank, and are therefore unable to change the ranking of the most central airports, which have large centrality gaps.

 \section{Elimination of new walks created by delays in the realised network}
 When computing Trip Centrality metrics in the realised network, new walks that were not present in the scheduled network appear due to the changes in the link schedules (i.e., in the air traffic case, delays). As explained in section 4.5 of the main text, in the air traffic case these walks should be excluded from the computation, as passengers cannot use them. On the contrary, for other transportation systems, e.g. bus and metro, itineraries do not need to be fixed in advance, therefore the new walks can be used. \\
 In this section, we show how we can exclude from the computation the new walks appearing when the delay of a departing flight allows the connection with an incoming flight, which was originally landing too late to make the connection. \\
 Consider Trip Centrality, and, as done in section 4.4 of the main text, let
\begin{equation}
    Q=(\mathbb{I}+\tilde{\alpha}A^{[1]}K)\dots(\mathbb{I}+\tilde{\alpha}A^{[T]}K).
\end{equation}
With this definition,
\begin{subequations}
\begin{align}
 \vec{t}^{out}&=[Q-\mathbb{I}]K^{-1}\vec{1}_{N N_L+N_l}, \\ 
 \vec{t}^{in}&=\vec{1}_{N N_L+N_l}^T[Q-\mathbb{I}]K^{-1}.
\end{align}
\end{subequations}
The element $Q_{ij}$ contains the contribution to the outgoing centrality of $i$ given by walks from $i$ to $j$ (or equivalently the contribution to the incoming centrality of $j$ given by walks from $i$ to $j$ ). We call $Q_{}$ the matrix computed with $A_{sched}$, and $Q_{r}$ the matrix computed with $A_{real}$ . Now, let us compute the matrix $Q$ one time frame at a time. $Q^{[1]}=(\mathbb{I}+A^{[1]}K)$ only counts the walks during the first time frame, $Q^{[2]}=(\mathbb{I}+A^{[1]}K)(\mathbb{I}+A^{[2]}K)$ counts the walks up to the second time frame, and so on. At each time step, if an element of $Q_{r}$ is larger than the corresponding element of $Q_{s}$, it is because of a new walk opened up by a delay. In fact, in the real network departure and landings take place either at the same time or later than in the scheduled network (having put all negative delays to zero, as motivated in section 4.5 of the main text), therefore all new acceptable contributions to centrality are added to $Q_{r}$ either at the same time as in $Q_{s}$, or later. Therefore, at each step we pose $Q_{r}=\min\{Q_{s},Q_{r}\}$. This eliminates most spurious walks. The correction procedure is analogous for TripRank, with the appropriate definition of $Q$.\\
In order to better understand the principle of this correction, let us consider a concrete example. Let $i$, $j$ and $k$ be three primary nodes, and let there be two flights from $i$ to $j$, f1 departing at $t=1$ and landing at $t=2$, and f2 departing at $t=3$ and landing at $t=4$. Additionally, assume there is a flight from $j$ to $k$, called f3, departing at $t=3$ and landing at $t=4$. According to schedule, it is possible to go from $i$ to $k$ with f1+f3, but not with f2+f3. However, assume now that f3 is delayed, and departs at $t=5$, arriving at $t=6$. In the realised network, it would now be possible to reach $j$ from $i$ also with f2+f3. However, the proposed correction eliminates this contribution. In fact, calling $a$ the secondary node associated with flight f3, we have $Q^{[5]}_{s, ia}=\tilde{\alpha}^3$, which counts the walk given by f1 and f3, and $Q^{[5]}_{r, ia}=2 \tilde{\alpha}^3$, which counts also the walk given by f2+f3. Now, taking $Q^{[5]}_{r, ia}=\min \{Q^{[5]}_{r, ia}, Q^{[5]}_{s, ia} \}= \tilde{\alpha}^3$, the combination f2+f3 is correctly removed. At the following timestep, then, we get $Q^{[6]}_{s, ik}=Q^{[6]}_{r, ik}=\tilde{\alpha}^4$. \\
We remark that there are still some `forbidden' walks that remain, even after this correction, however they are very rare. Specifically, if in the example above f1 was delayed such that the combination f1+f3 became impossible, while the delay of f3 still makes the combination f2+f3 possible in the realised network, in the realised network we should obtain that $Q^{[6]}_{r, ik}=0$, as one walk is impossible and the other one is forbidden. However, at $t=5$ we have $Q^{[5]}_{s, ia}=\tilde{\alpha}^3$, which counts the walk given by f1 and f3, and $Q^{[5]}_{r, ia}= \tilde{\alpha}^3$, which counts only the walk given by f2+f3. Then, at the following time step we would get $Q^{[6]}_{s, ik}=Q^{[6]}_{r, ik}=\tilde{\alpha}^4$. Therefore, in this case, a walk of two flights is still counted in the realised network, although it is forbidden. However, the realisation of this situation requires an improbable combination of delays. For example, on April 1st there are only 142 such combinations, on a total of almost 2 millions triplets of flights that could potentially give rise to such combinations. Additionally,  82 of these 142 are inter-layer, therefore not counted when $\varepsilon=0$. We conclude that the number of forbidden walks remaining after the correction introduced in this section is so small that they give a negligible contribution to centrality and therefore do not invalidate the comparison between the scheduled and realised networks. 
 
 \newpage
 
 \section{Supplementary tables and figures}
 
\begin{figure}[h!]
\begin{center}
\includegraphics{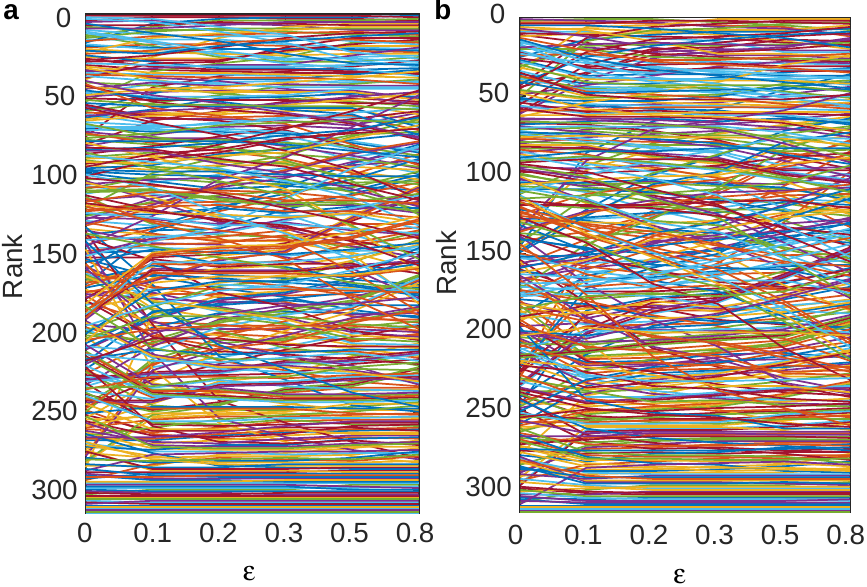}
\caption{Evolution of the airports' ranking according to outgoing (a) and incoming (b) Trip Centrality on the scheduled network on April 1st for $\alpha=0.2$ and different values of $\varepsilon$. Each line represents one airport, and the position on the $y$-axis indicates its rank for each $\varepsilon$ value. Rank 1 corresponds to the most central airport. }
\label{fig:comparison_e}
\end{center}
\end{figure}

\begin{figure}[h!]
\begin{center}
\makebox[\textwidth][c]{\includegraphics{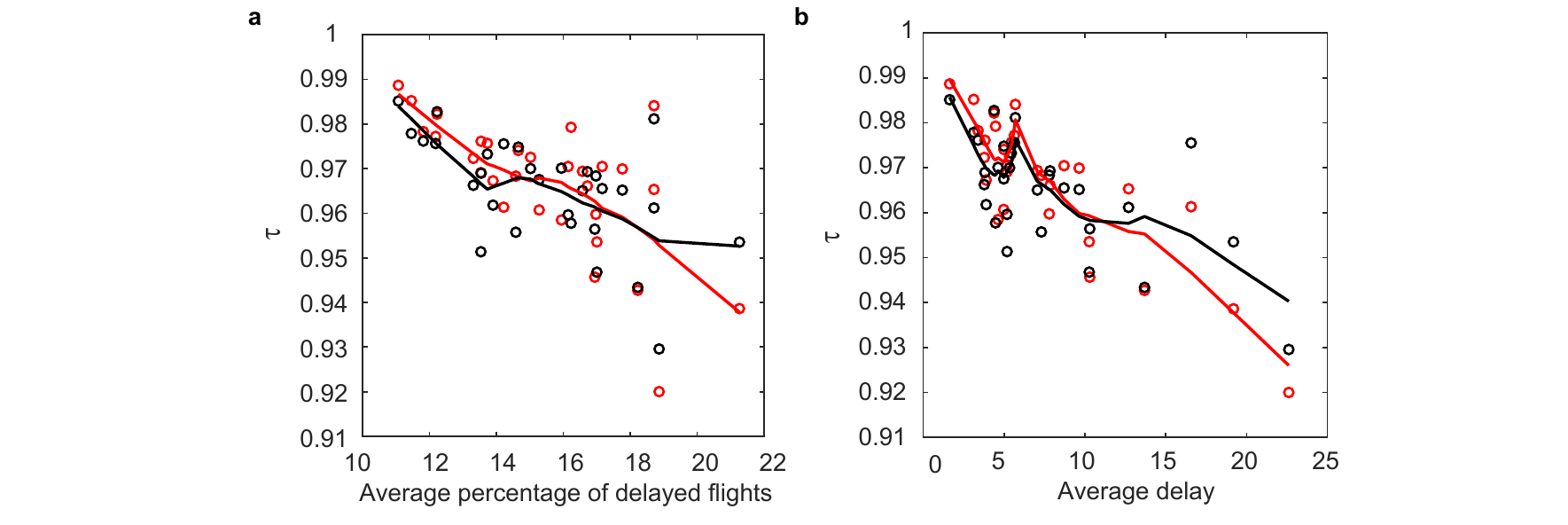}}%
\caption{Kendall correlation coefficient between the airports' ranking on the scheduled and realised network, in each day of the dataset, according to incoming Trip Centrality (red) and outgoing Trip Centrality (black), plotted against average departure delay (a) and average fraction of flights with departure delay in one airport (b). Each point corresponds to one day of the dataset. Lines are obtained by a locally weighted smoothing (LOWESS) of the dots of the correspondent color.}
\label{fig:av_rank_corr}
\end{center}
\end{figure}

\begin{figure}[h!]
\begin{center}
\includegraphics{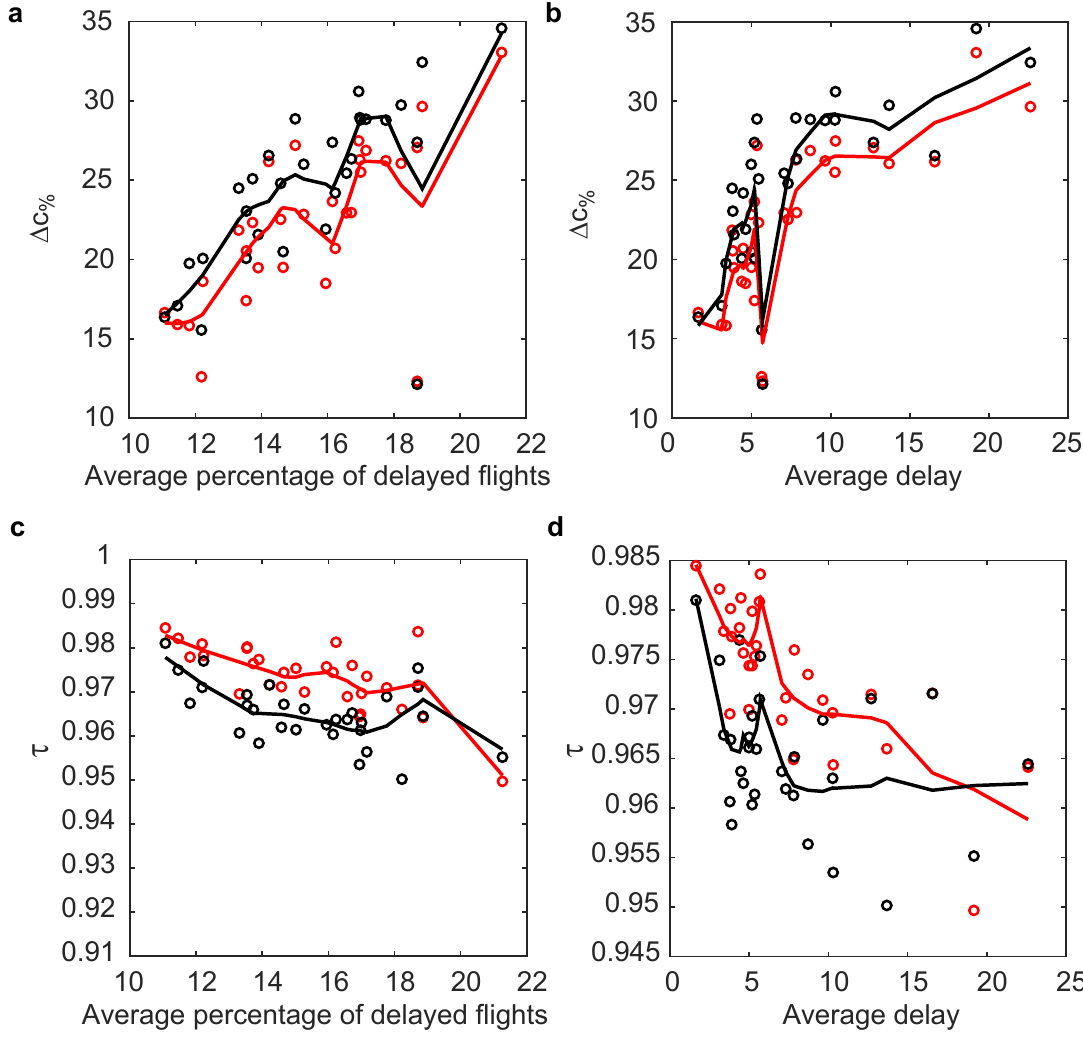}
\caption{Replica of the analysis in section 2.3.3 of the main text, excluding cancelled and diverted flights. a and b: Percentage of centrality loss, averaged over all airports, in each day of the dataset, plotted against the average percentage of delayed flights in an airport (a) and the average delay of flights (b), according to incoming Trip Centrality (red circles) and outgoing Trip Centrality (black circles).  Trip centrality is computed with $\alpha=0.2$ and $\varepsilon=0$. The percentage centrality loss of an airport is computed as $\Delta c_{\%}= 100 \times (c_{sched}-c_{act})/c_{sched}$, where $c_{sched}$ and $c_{act}$ are the airport's centralities on the scheduled and realized network. Lines are obtained by a locally weighted smoothing (LOWESS) of the circles of the correspondent color. c and d: Kendall correlation coefficient between the airports' ranking on the scheduled and realised network plotted against the average percentage of delayed flights in an airport (c) and the average delay of flights (d), according to incoming Trip Centrality (red circles) and outgoing Trip Centrality (black circles). Lines are obtained by a locally weighted smoothing (LOWESS) of the circles of the correspondent color.}
\label{fig:no_canc_tc}
\end{center}
\end{figure}

\begin{figure}[h!]
\begin{center}
\includegraphics{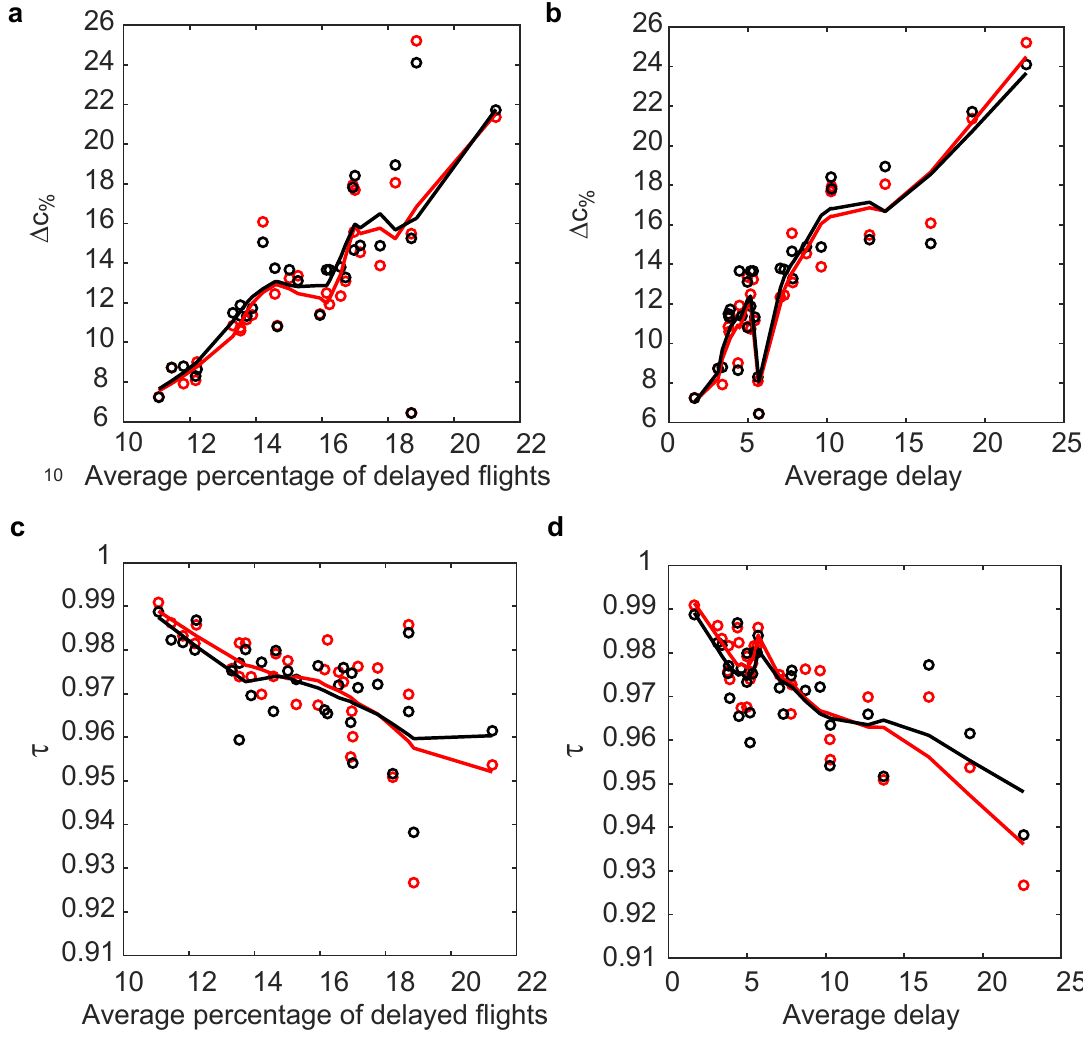}
\caption{Replica of the analysis in section 2.3.3 of the main text with $\alpha=0.1$. a and b: Percentage of centrality loss, averaged over all airports, in each day of the dataset, plotted against the average percentage of delayed flights in an airport (a) and the average delay of flights (b), according to incoming Trip Centrality (red circles) and outgoing Trip Centrality (black circles).  The percentage centrality loss of an airport is computed as $\delta c_{\%}= 100 \times (c_{sched}-c_{act})/c_{sched}$, where $c_{sched}$ and $c_{act}$ are the airport's centralities on the scheduled and realized network. Lines are obtained by a locally weighted smoothing (LOWESS) of the circles of the correspondent color. c and d: Kendall correlation coefficient between the airports' ranking on the scheduled and realised network plotted against the average percentage of delayed flights in an airport (c) and the average delay of flights (d), according to incoming Trip Centrality (red circles) and outgoing Trip Centrality (black circles). Lines are obtained by a locally weighted smoothing (LOWESS) of the circles of the correspondent color.}
\label{fig:robcheck_01}
\end{center}
\end{figure}

\begin{figure}[h!]
\begin{center}
\includegraphics{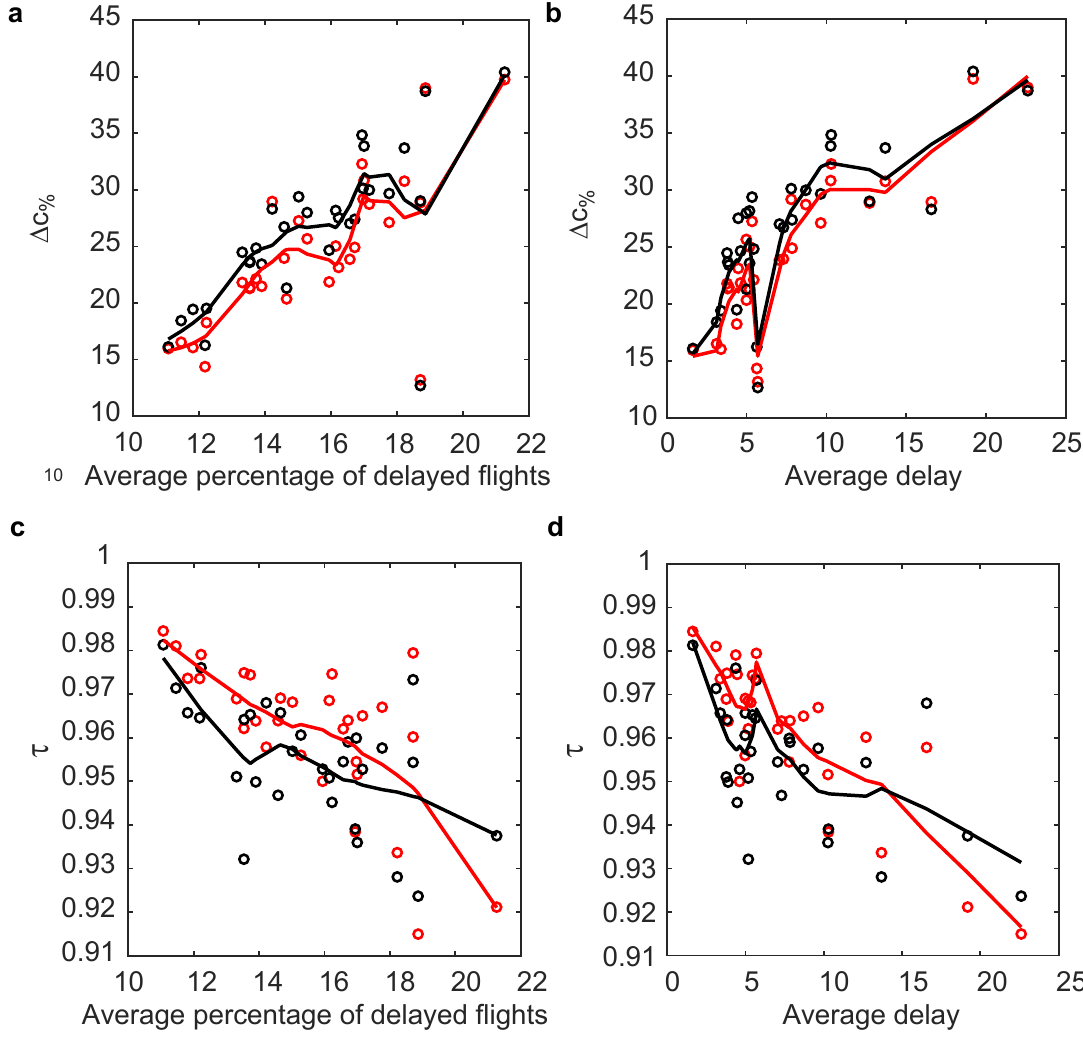}
\caption{Replica of the analysis in section 2.3.3 of the main text with $\alpha=0.5$. a and b: Percentage of centrality loss, averaged over all airports, in each day of the dataset, plotted against the average percentage of delayed flights in an airport (a) and the average delay of flights (b), according to incoming Trip Centrality (red circles) and outgoing Trip Centrality (black circles).  The percentage centrality loss of an airport is computed as $\delta c_{\%}= 100 \times (c_{sched}-c_{act})/c_{sched}$, where $c_{sched}$ and $c_{act}$ are the airport's centralities on the scheduled and realized network. Lines are obtained by a locally weighted smoothing (LOWESS) of the circles of the correspondent color. c and d: Kendall correlation coefficient between the airports' ranking on the scheduled and realised network plotted against the average percentage of delayed flights in an airport (c) and the average delay of flights (d), according to incoming Trip Centrality (red circles) and outgoing Trip Centrality (black circles). Lines are obtained by a locally weighted smoothing (LOWESS) of the circles of the correspondent color.}
\label{fig:robcheck_05}
\end{center}
\end{figure}

\begin{figure}[h!]
\begin{center}
\includegraphics{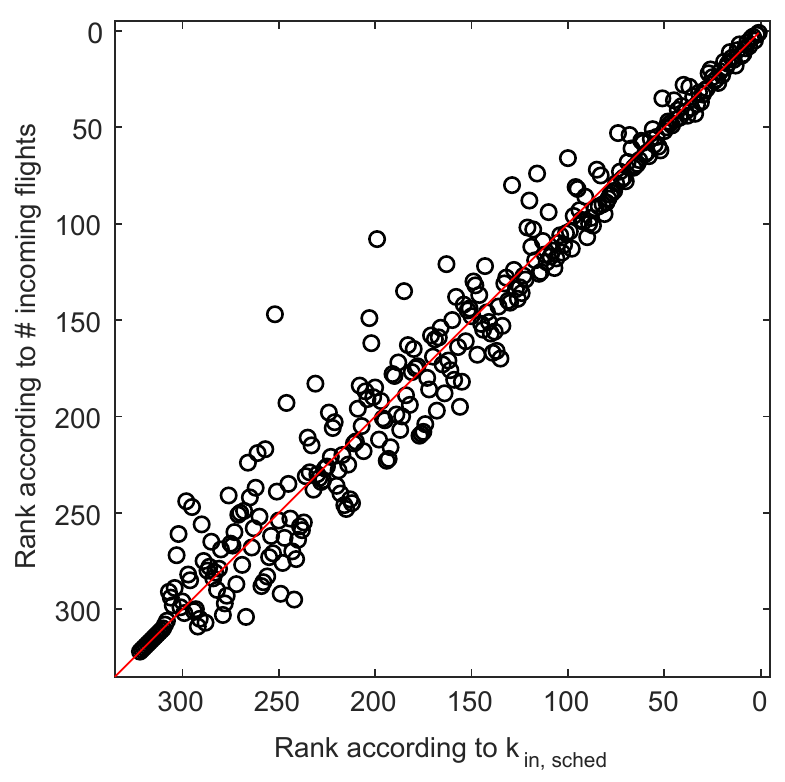}
\caption{Comparison of the ranking according to incoming Katz centrality with $\alpha=0.003$ and the ranking according to the number of incoming flights for April 1st, on the scheduled network. }
\label{fig:katzVSstrenght}
\end{center}
\end{figure}

\begin{figure}[t]
\begin{center}
\makebox[\textwidth][c]{\includegraphics{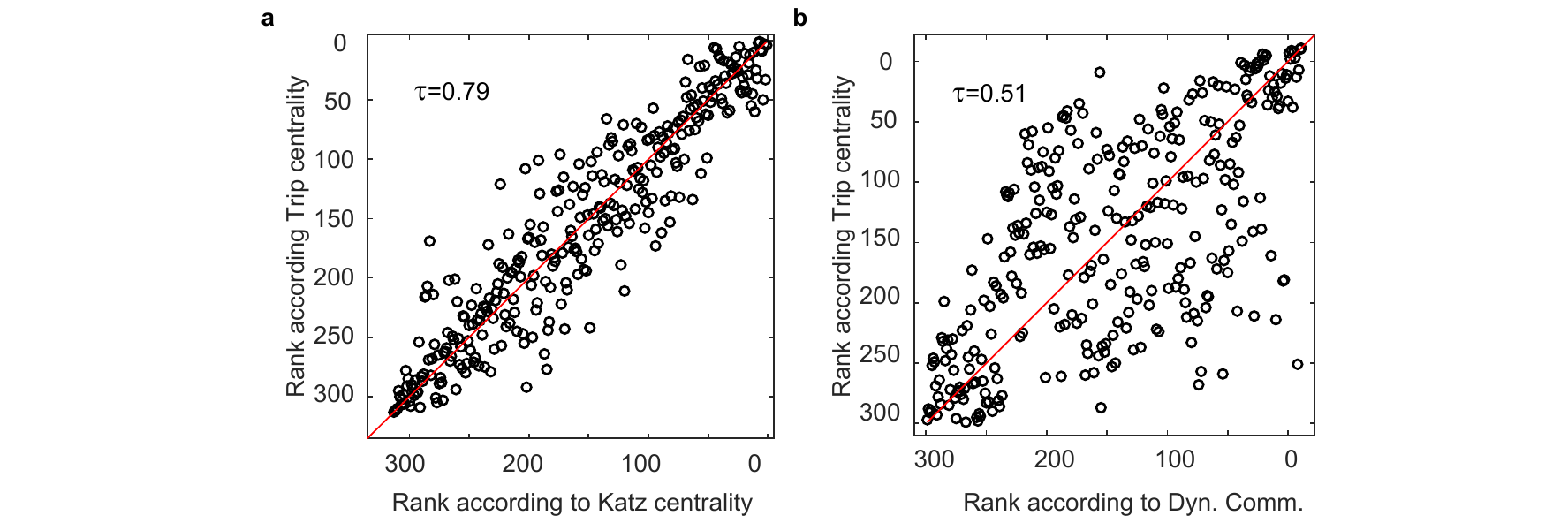}}%
\caption{Comparison of airports' rankings on April 1st according to a) incoming  Katz centrality and Trip Centrality and b) incoming Dynamic communicability and Trip Centrality. Trip Centrality is computed with $\alpha=0.2$ and $\varepsilon=0$. Each circle represents an airport, rank 1 corresponds to the highest centrality. The red line is the 1:1 line. The Kendall correlation coefficients are reported in the figure.}
\label{fig:rank_comp_stat_trip}
\end{center}
\end{figure}

\begin{table}[h!]
 \label{tab:topten}
 \begin{center}
 \begin{small}
\begin{tabular}{lll}

 \textbf{Katz centrality} & \textbf{Trip Centrality} &\textbf{Dynamic communicability} \\
 \hline
 Atlanta	& Phoenix  & Chicago O'Hare \\
Chicago O'Hare & Las Vegas McCarran & New York (JFK) \\
Los Angeles & Los Angeles  & Phoenix \\
Dallas/Fort Worth  & Atlanta & Boston Logan \\
Denver  & San Diego & Miami  \\
San Francisco & Oakland  & Charlotte Douglas \\
Phoenix  & Sacramento  & Denver \\
Las Vegas McCarran  & San Francisco  & Philadelphia  \\
LaGuardia  & Denver  & Los Angeles \\
Boston Logan  & Orlando & Newark  

\end{tabular}
\end{small}
\caption{Top ten airports on April 1st according to incoming Katz Centrality, Trip Centrality and Dynamic communicability. Katz centrality is computed with $\alpha=0.003$, Trip Centrality and Dynamic communicability with $\alpha=0.2$, Trip Centrality is computed with $\varepsilon=0$. }
\end{center}
\end{table}

\begin{figure}[h!]
\begin{center}
\makebox[\textwidth][c]{\includegraphics{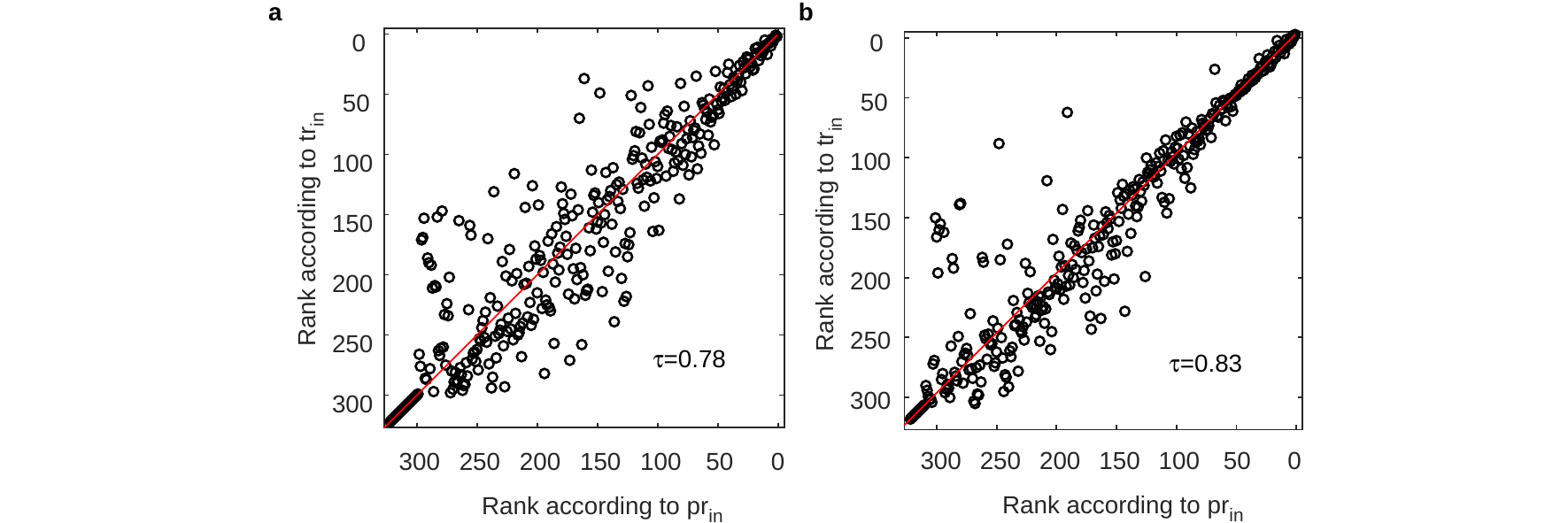}}%
\caption{Comparison of the rankings according to incoming TripRank  and PageRank for April 1st, on the scheduled network. TripRank is computed with $\alpha=0.2$ and with $\varepsilon=0$ in panel a) and $\varepsilon=0.8$ in panel b). The red line is the 1:1 line.}
\label{fig:prVStr}
\end{center}
\end{figure}

\begin{figure}[h!]
\begin{center}
\includegraphics{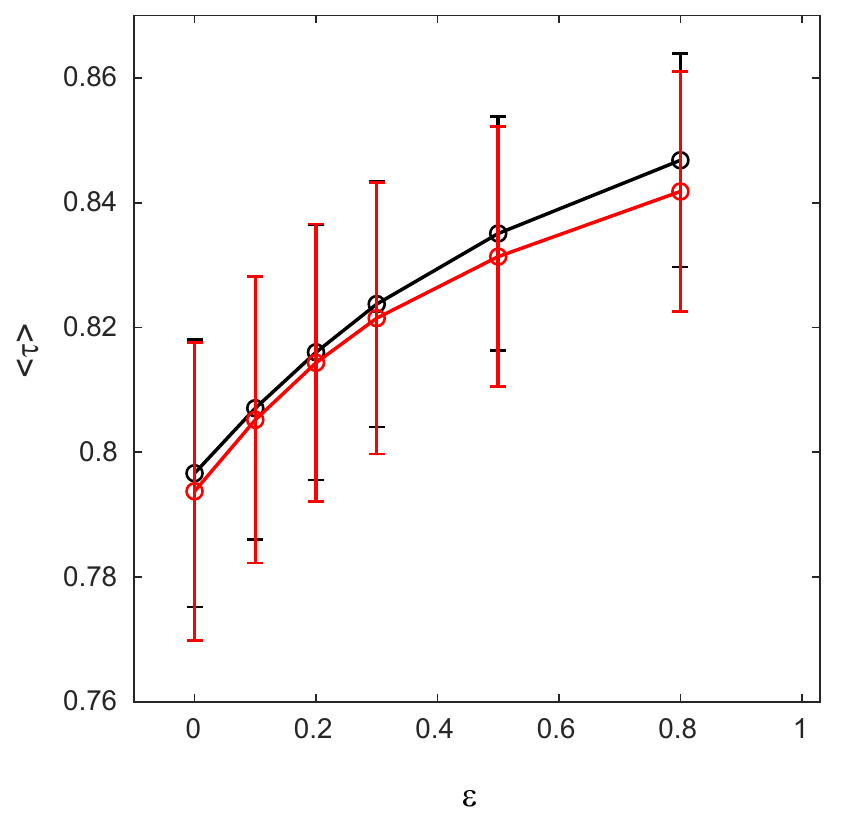}
\caption{Kendall correlation coefficient, averaged on all days, between the rankings according to TripRank and Page Rank, for different values of $\varepsilon$ used in the computation of TripRank. TripRank is computed with $\alpha=0.2$. The red line corresponds to the incoming centralities and the black one to the outgoing. Error bars represent standard errors. }
\label{fig:prVStr_corr}
\end{center}
\end{figure}

\begin{table}[h!]
\label{tab:topten_TR}

\begin{center}
\begin{small}
\begin{tabular}{lll}

\textbf{PageRank}  & \textbf{TripRank ($\varepsilon=0$)} & \textbf{TripRank ($\varepsilon=0.3$)}    \\
 \hline
Atlanta & Atlanta & Atlanta\\
Chicago O'Hare & Dallas/Fort Worth& Dallas/Fort Worth\\
Dallas/Fort Worth  &  Chicago O'Hare &  Chicago O'Hare\\
Los Angeles & Denver & Denver\\
Denver  & George Bush & George Bush\\
Phoenix  & Salt Lake City  &  Salt Lake City\\
San Francisco  &  Los Angeles & Los Angeles\\
George Bush  & Minneapolis-Saint Paul &  San Francisco\\
Las Vegas McCarran  & San Francisco & Minneapolis-Saint Paul\\
Boston Logan  & Detroit & Detroit

\end{tabular}
\end{small}
\end{center}
\caption{Top ten airports on April 1st according to incoming PageRank and TripRank. TripRank is computed with $\alpha=0.2$.}
\end{table}

\clearpage 
\bibliographystyle{naturemag}
\bibliography{TripCentrality.bib}